\documentclass[fleqn,usenatbib]{mnras}

\usepackage{newtxtext,newtxmath}

\usepackage[T1]{fontenc}
\usepackage{ae,aecompl}

\usepackage{graphicx}	
\usepackage{amsmath}	
\usepackage{amssymb}	

\title[the SLSN-GRB Connection]{The GRB-SLSN Connection: mis-aligned magnetars, weak jet emergence, and observational signatures}

\author[B. Margalit {\it et al.}]{Ben Margalit,$^{1}$\thanks{E-mail: 
\href{mailto:btm2134@columbia.edu}{btm2134@columbia.edu}}
Brian D.~Metzger,$^{1}$ 
Todd A.~Thompson,$^{2}$ 
Matt Nicholl$^{3}$ 
and
\newauthor
Tuguldur Sukhbold$^{2}$
\\
$^{1}$Columbia Astrophysics Laboratory, Columbia University, New York, NY 10027, USA\\
$^{2}$Department of Astronomy and Center for Cosmology \& Astro-Particle Physics, The Ohio State University, Columbus, OH 43210, USA\\
$^{3}$Harvard-Smithsonian Center for Astrophysics, 60 Garden Street, Cambridge, MA 02138, USA
}

\date{Accepted XXX. Received YYY; in original form ZZZ}

\pubyear{2017}

\begin{document}
\label{firstpage}
\pagerange{\pageref{firstpage}--\pageref{lastpage}}
\maketitle

\begin{abstract}
Multiple observational lines of evidence support a connection between hydrogen-poor superluminous supernovae (SLSNe) and long duration gamma-ray bursts (GRBs). Both events require a powerful central energy source, usually attributed to a millisecond magnetar or an accreting black hole. The GRB-SLSN link raises several theoretical questions: What distinguishes the engines responsible for these different phenomena? Can a single engine power both a GRB and a luminous SN in the same event? We propose a new unifying model for magnetar thermalization and jet formation: misalignment between the rotation (${\bf \Omega}$) and magnetic dipole (${\bf \mu}$) axes thermalizes a fraction of the spindown power by reconnection in the striped equatorial wind, providing a guaranteed source of ``thermal" emission to power the supernova. The remaining un-thermalized power energizes a relativistic jet. In this picture, the GRB-SLSN dichotomy is directly linked to ${\bf \Omega \cdot \mu}$
through the misalignment angle $\alpha$, such that the thermal fraction is $\simeq 1.025 \alpha / \left( 0.636 + \alpha^4 \right)^{1/4}$. 
We extend earlier work to show that even weak relativistic jets of luminosity $\sim10^{46}$ erg s$^{-1}$ can escape the expanding SN ejecta hours after the explosion --- whether driven by a magnetar or black hole accretion --- implying that escaping relativistic jets may accompany many SLSNe regardless of the details of their driving mechanism.
We calculate the observational signature of these jets. We show that they may produce transient UV cocoon emission lasting a few hours when the jet breaks out of the ejecta surface. A longer-lived optical/UV signal may originate from a mildly-relativistic wind driven from the interface between the jet and the ejecta walls. This provides a new explanation for the secondary early-time maximum observed in some SLSNe light curves, such as LSQ14bdq. This scenario also predicts a population of GRB from on-axis jets with extremely long durations, potentially similar to the population of ``jetted tidal disruption events", in coincidence with a small subset of SLSNe.
\end{abstract}



\section{Introduction}

Type I ``superluminous supernovae" (SLSNe; \citealt{Quimby+11}) are a rare class of core collapse supernovae (SNe) with hydrogen-poor spectra that reach peak luminosities $\sim 10^{43}-10^{45} \, {\rm erg \, s}^{-1}$ which are usually too large to be powered by the radioactive decay of $^{56}$Ni.  The nickel mass required to power these events would in most cases exceed the total ejecta mass inferred from the timescale of the light curve rise (e.g. \citealt{Inserra+13,Nicholl+13}; however see \citealt{Kozyreva+17}); the UV-rich spectrum of SLSNe also disfavors the ejecta being so rich in iron-group elements (\citealt{Kasen+11,Dessart+12}).  These facts argue against most SLSNe being the result of pair instability in extremely massive stars  (\citealt{Barkat+67,Kasen+11,Dessart+13,Kozyreva+17}).\footnote{Pair instability supernovae could remain a contender for explaining slowly-evolving SLSNe (e.g., \citealt{GalYam+09,Lunnan+16}) if the persistent blue colors can somehow be reconciled with the predicted high abundance of Fe-group elements in the ejecta.}

Another potential power source for SLSNe is a shock-mediated collision between the SN ejecta and the external circumstellar medium (\citealt{Chevalier&Irwin11,Ginzburg&Balberg12,Moriya+13,Chatzopoulos+13}).  The latter might be produced by pre-explosion stellar mass loss (as in Type IIn supernovae; \citealt{Chevalier&Fransson94,Moriya+14,Dessart+15}), pulsational pair instability 
(e.g.~\citealt{Woosley+07,Chatzopoulos&Wheeler12,Woosley16}), or a relic proto-stellar disk 
(\citealt{Metzger10}).  However, the lack of spectroscopic evidence such as emission lines for shock interaction or the presence of extended circumstellar matter again appears to disfavor interaction models for the majority of Type I SLSNe \citep{Nicholl+15}.  Line-emitting gas could be sufficiently deeply embedded to remain hidden at all times, but this appears to require fine-tuning of the distribution of external matter. 
A small number of SLSNe-I show clear signatures of interaction through late-time H$\alpha$ emission \citep{Yan+15,Yan+17}, but these occur at late, $\gtrsim 100 \,{\rm d}$, epochs indicating that such interaction does not power the peak luminosity.

The most strongly favored model for SLSNe is sustained energy input from the young compact stellar remnant, such as the electromagnetic dipole spin-down of a strongly magnetized neutron star (``millisecond magnetar''; \citealt{Kasen&Bildsten10,Woosley10,Metzger+14}) or an accreting black hole \citep{Dexter&Kasen13}.  The simplest form of the magnetar model --- which assumes that 100\% of the spin-down power of the magnetar is converted to thermal energy behind the ejecta --- predicts a light curve evolution which can be successfully fit to those of most SLSNe \citep{Inserra+13,Chatzopoulos+13,Nicholl+14}.  

The association between SLSNe and the birth of energetic compact objects is further supported by the growing observational connection between SLSNe and long duration gamma-ray bursts (GRB).  GRB originate from the deaths of a rare class of massive stars, a fact well-established by their observed coincidence with hyper-energetic, broad-lined Type Ic SNe (SN Ic-bl; e.g.~\citealt{Stanek+03,Hjorth+03,Woosley&Bloom06,Modjaz+16}).  As in SLSNe, the relativistic jets responsible for powering GRB could be fed by the rotational energy of a millisecond magnetar (e.g., \citealt{Usov92,Wheeler+00,Thompson+04,Metzger+11}) or the accretion energy of a stellar mass black hole (e.g.~\citealt{MacFadyen&Woosley99,MacFadyen+01}).  SLSNe and SNe Ic-bl have comparable average spectral features and absorption line velocities at all phases, which are systematically broader than those of normal SNe Ic \citep{Liu&Modjaz16}, while also sharing remarkably similar nebular spectra \citep{Nicholl+16b,Jerkstrand+17}.  Both types of events exhibit a strong preference to occur in low-luminosity, metal-poor star-forming galaxies (e.g.~\citealt{Stanek+06,Chen+15,Perley+16a, Japelj+16, Perley+16b}).  This shared rare environment (compared to where most star formation in the universe occurs) hints at a common progenitor to both phenomena, likely associated with whatever special conditions are needed to produce presupernova stellar cores with extremely high rotation.  

Further indirect evidence for a connection between SLSNe, GRB, and magnetar birth was provided by the recent localization of the repeating fast radio burst (FRB) FRB 121102 \citep{Chatterjee+17} in a dwarf irregular galaxy \citep{Tendulkar+17} with properties of mass, metallicity, and star formation remarkably similar to the hosts of SLSNe and GRB \citep{Tendulkar+17,Metzger+17,Nicholl+17b}.  Prior to this discovery, several works hypothesized that FRBs originate from flaring magnetars (e.g.~\citealt{Lyubarsky14,Kulkarni+14,Lyutikov+16}), thus supporting an association between FRB 121102 and the birth of a millisecond magnetar embedded in a young supernova remnant with an age of decades to a century \citep{Metzger+17,Beloborodov17,Lyutikov17,Waxman17}. 

If both SLSNe and GRB are engine-powered, it is natural to question what distinguishes them.  Their most important distinction is probably the {\it duration} of the engine's peak luminosity, which is characteristically minutes or less in GRB, and usually days or longer in the case of SLSNe (e.g.~\citealt{Metzger+15}).  A long-lived engine is essential in SLSNe to enhance their luminosity above the minimum level set by radioactive $^{56}$Ni, the latter of which instead powers the luminosity of the Ic-bl SNe accompanying most GRB (e.g.~\citealt{Cano+16}; however, see \citealt{Wang+16}).  

A class of ultra-long GRB (ULGRB) with durations of $\gtrsim 10^3 \, {\rm s}$ has recently received attention \citep{Gendre+13,Levan+14,Zhang+14,Boer+15}.  
The ULGRB 111209A was observed in coincidence with a highly luminous $\sim 10^{43} \, {\rm erg \, s}^{-1}$ and short-lived $\sim 15 \, {\rm d}$ supernova \citep{Greiner+15} with a blue spectrum consistent with those associated with SLSNe (\citealt{Liu&Modjaz16,Kann+16}), providing a potential direct link between GRB and SLSNe \citep{Metzger+15,Bersten+16,Gompertz&Fruchter17}.  In retrospect, this connection is unsurprising because a long-lived engine (as required to power an ultra-long GRB) is identical to that needed to maintain a luminosity $\sim 10^{43} \, {\rm erg \, s}^{-1}$ a couple weeks after the explosion, near the super-luminous level \citep{Metzger+15}.  

On the other hand, ULGRB 111209A raises the question of how it is possible theoretically for a single engine to power both a successful jet (which escapes the exploding star to power the GRB) while also thermalizing a large fraction of its energy behind the ejecta (to power the SN at later times).  If a jet can escape from the ejecta of a SLSNe, then it is also natural to ask what observable signatures such jets would display for the more typical off-axis observer.

The key questions raised above motivate additional theoretical studies of the GRB-SLSN connection and ways to test it observationally. \citet{Metzger+15} recently highlighted the diversity of transients potentially associated with millisecond magnetar birth, providing a common theoretical framework for interpreting diverse phenomena from long-duration GRB (LGRB) and ULGRB to SLSNe and broad-line SN-Ic (see also \citealt{Kashiyama+16,Ioka+16}). 

This work goes beyond these initial steps to describe an explicit mechanism by which magnetars can power both a GRB jet and a SLSN. 
In particular, previous analytic and numerical calculations either explicitly or implicitly assumed ${\bf \hat{\Omega} \cdot \hat{\mu}}=1$. 
In general, we expect mis-alignment between the magnetic and rotation axes, implying an explicit mechanism for thermalization of the spindown power via reconnection in the equatorial striped magnetar wind \citep{Lyubarsky03}. 
For a range of mis-alignment angles we expect either complete thermalization 
(${\bf \hat{\Omega} \cdot \hat{\mu}}=0)$, as assumed in models of SLSNe \citep{Kasen&Bildsten10,Woosley10}, or virtually no thermalization and strong jet production 
(${\bf \hat{\Omega} \cdot \hat{\mu}}=1)$, as in models of GRBs (e.g., \citealt{Bucciantini+09}).

We develop this model and apply it. For the weak jets that we expect are launched generically by SLSNe, we extend earlier work to ask whether or not they can escape the supernova explosion, on what timescale, and with what observational signature, whether the observer is on- or off-axis. Our general model of thermalization and jet production allows us to provide a unified picture of the GRB-SLSN dichotomy and connection. 

Setting the details of the magnetar thermalization mechanism we propose aside, our estimates for low-luminosity jet emergence and its observational signature can also be applied to black hole accretion models.

This paper is organized as follows.  
In  \S\ref{sec:MagnetarModel} we review the magnetar scenario and present our model for partitioning spin-down luminosity between both jetted and thermal components. 
We then examine whether weak jets can break-out of the confining stellar matter (\S\ref{sec:Jet}). Readers uninterested in the jet-propagation details are encouraged to skip to \S\ref{subsubsec:breakout_from_ejecta} where we derive our primary results.
We continue by exploring the observational signatures such off-axis jets may give rise to (\S\ref{subsec:JetSignature}). Our novel model for powering early optical/UV emission in SLSNe by (post-breakout) jet interaction with the confining SN-ejecta walls is presented in \S\ref{subsubsec:jet_energized_emission} and applied to the SLSN LSQ14bdq.
We discuss implications of our results in \S\ref{sec:Discussion} and summarize the landscape of engine-powered transients in Figure \ref{fig:ParameterSpace}. We end with bulleted conclusions (\S\ref{sec:Conclusions}).

\section{Magnetar Misalignment: Powering Both Jet and SN} \label{sec:MagnetarModel}
\begin{figure}
\centering
\includegraphics[width=0.45\textwidth]{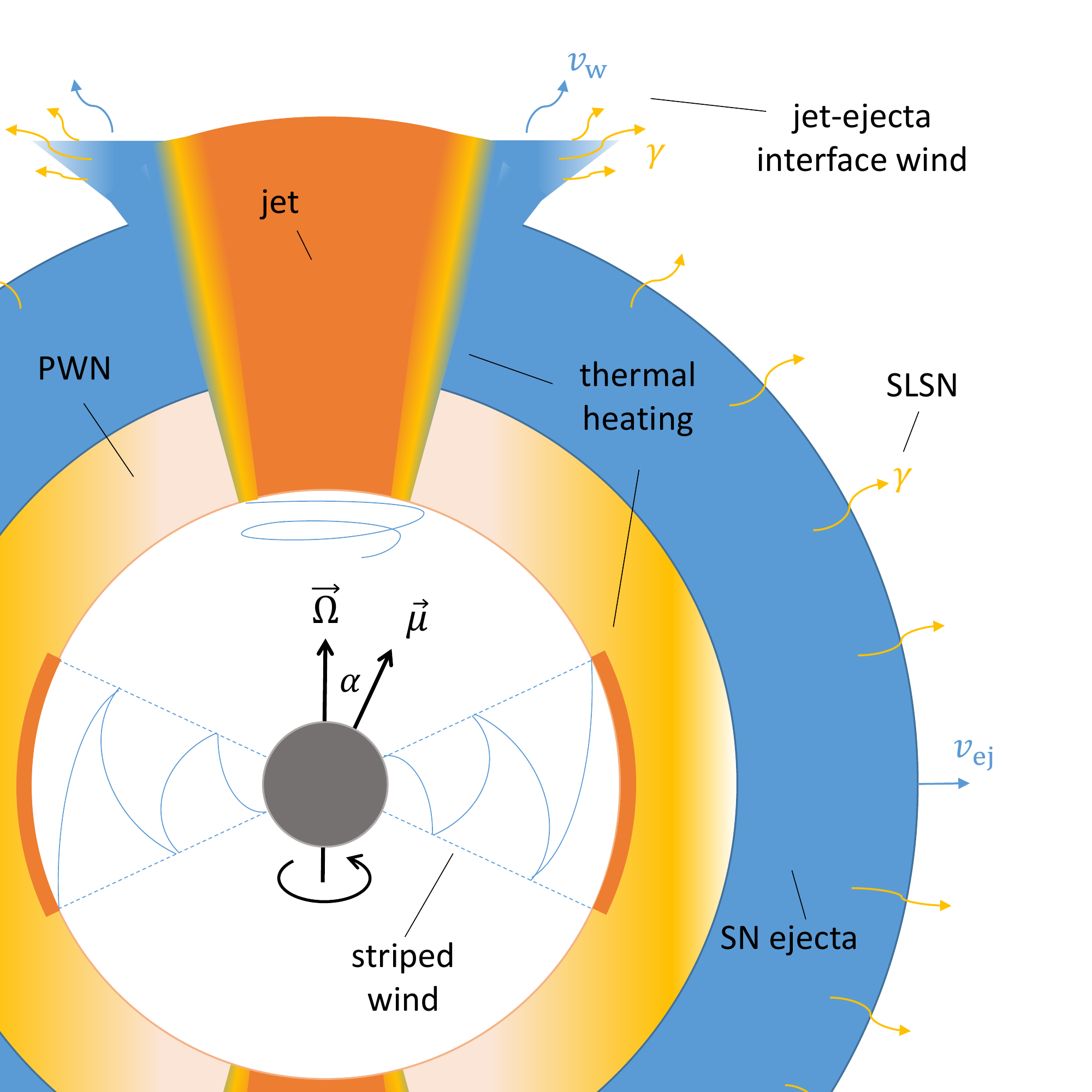}
\caption{Schematic diagram (not to scale) showing how the same millisecond magnetar engine can power both a relativistic GRB jet and a SLSN via isotropic radiative diffusion.  A magnetar (grey) with a non-zero misalignment between the rotation and magnetic dipole axes develops a striped-wind configuration in a wedge near the equatorial plane.  The fraction of the spin-down energy carried by the striped wind is thermalized when the alternating field undergoes magnetic reconnection near the wind termination-shock, 
heating the pulsar-wind nebula (PWN; yellow). This thermal energy diffuses through the spherical SN ejecta (blue), powering luminous SN emission.  By contrast, the spin-down power at high latitudes is channeled into a bi-polar collimated jet (orange; \S\ref{sec:MagnetarModel}).  Even once the jet has escaped from the star, a fraction of its power will continue to be thermalized at the interface between the jet and the ejecta walls, driving a hot mildly-relativistic wind of velocity $v_{\rm w}$.  Thermal radiation from this wind may give rise to relatively isotropic optical/UV emission viewable off the jet axis, producing a pre-maximum peak in the light curves of SLSNe  (\S\ref{subsubsec:jet_energized_emission}; Fig.~\ref{fig:lightcurve}).
} \label{fig:Cartoon}
\end{figure}

We begin this section with a brief review of the magnetar model, following which we describe an explicit mechanism by which a misaligned magnetar can partition its power into both thermal and magnetically-dominated (jetted) components.
The engine luminosity's time evolution can generally be expressed as \citep{Kasen&Bildsten10}
\begin{equation}
L_{\rm e} = \frac{E_{\rm e}}{ t_{\rm e}} \frac{(\ell-1)}{\left( 1 + t/t_{\rm e} \right)^\ell} ,
\label{eq:Le}
\end{equation}
where $E_{\rm e}$ is the total energy of the engine and $t_{\rm e}$ the engine lifetime, over which the power is approximately constant. At late times $t \gg t_{\rm e}$ the power decays as $\sim t^{-\ell}$. 
For the magnetar scenario, $\ell=2$ and the values of $E_{\rm e}$ and $t_{\rm e}$ are related to the magnetar's surface dipole field\footnote{As in \cite{Metzger+15}, our definition of $B_{\rm d}$ is a factor of $\sqrt{12}$ lower than the normalization adopted by \cite{Kasen&Bildsten10} (as commonly adopted in the SLSNe literature).}, $B_{\rm d}$, and initial spin period, $P_0$, according to
\begin{equation} \label{eq:Ee_magnetar}
E_{\rm e} = \frac{1}{2} I_{\rm ns} \Omega^2 \simeq 2.5 \times 10^{52} \, {\rm erg} \left(\frac{M_{\rm ns}}{1.4 M_\odot}\right)^{3/2} \left(\frac{P_0}{1 \, {\rm ms}}\right)^{-2} ,
\end{equation}
\begin{align}  \label{eq:te_magnetar}
t_{\rm e} &= \frac{E_{\rm e} c^3}{\mu^2 \Omega^4 \left(1 + \sin^2 \alpha\right)} \\ \nonumber
&\simeq \frac{147 \, {\rm s}}{\left(1 + \sin^2 \alpha\right)} \left(\frac{M_{\rm ns}}{1.4 M_\odot}\right)^{3/2} 
\left(\frac{P_0}{1 \, {\rm ms}}\right)^{-2}
\left(\frac{B_{\rm d}}{10^{15} \, {\rm G}}\right)^{-2}, 
\end{align}
where $I_{\rm ns} \simeq 1.3 \times 10^{45} \, {\rm g \, cm}^2 (M_{\rm ns} / 1.4 M_\odot)^{3/2}$ is an estimate of the neutron-star moment of inertia for a range of plausible nuclear density equations of state \citep{Lattimer&Schutz05}, $\mu = B_{\rm d} R_{\rm ns}^3$ is the magnetic dipole moment, and the factor $(1 + \sin^2 \alpha)$ accounts for the dependence on the misalignment angle $\alpha$ between magnetic and rotational axes ($\cos \alpha \equiv \hat{\bf \Omega} \cdot \hat{\bf \mu}$, see Fig.~\ref{fig:Cartoon}; \citealt{Spitkovsky06}).  We assume that all of the rotational energy goes into electromagnetic spin-down, instead of gravitational wave radiation (e.g.~\citealt{Moriya&Tauris16,Ho16}).

The notion of simultaneously powering both a collimated jet and an isotropic thermal SN  by a single magnetar has previously been discussed, e.g. in the context of relating hyper-energetic broad-lined Ic SNe to GRBs \citep[e.g.][]{Thompson+04}.
Such models, though, could not address a fundamental question --- {\it how is the magnetar energy partitioned between jet and SN?}
Later numerical simulations by \cite{Bucciantini+09} found that nearly none of the magnetar spindown power was deposited into the spherical SN component \citep[see also][]{Komissarov&Barkov07}, raising questions as to the viability of magnetar-driven SNe.
Here, we propose a solution to this ``thermalization problem'' by introducing a new, explicit, model for magnetar thermalization.
The idea rests on consideration of the mis-alignment angle $\alpha$ between the magnetar's rotation and magnetic axes.
In this respect, the 2D axisymmetric simulations mentioned above implicitly assumed $\alpha=0$, and could not capture the physics of our proposed model.

For $\alpha \neq 0$, the magnetar develops a `striped-wind' configuration where the toroidal magnetic field switches polarity in the equatorial plane \citep{Coroniti90,Lyubarsky&Kirk01}, as illustrated schematically in Fig.~\ref{fig:Cartoon}. The consequences of this wind geometry are well-studied in the pulsar community, as they may play an role in solving the so-called ``$\sigma$ problem" first identified in the Crab Nebula.

Outside of the light cylinder, in the wind zone, the power pattern varies with latitude $\theta$ (measured from the rotation axis) as $dL_{\rm e}/d\Omega \propto \sin^2\theta$. For a misaligned rotator, a fraction of the magnetic energy at low latitudes (within $\pm \alpha$ from the equator) will be dissipated by forced reconnection of the striped wind in the equatorial wedge near the temination shock which separates the wind from the magnetar nebula. Following \cite{Lyubarsky03} and \cite{Komissarov13}, the fraction of the wind power remaining in Poynting flux at latitude $\theta$ is given by
\begin{equation}
\chi(\theta; \alpha) = 
\begin{cases}
1,	&0 \leq \theta < {\pi}/{2}-\alpha \\
\left[{2\phi(\theta;\alpha)}/{\pi} - 1\right]^2,	&{\pi}/{2}-\alpha \leq \theta < {\pi}/{2}
\end{cases} ,
\end{equation}
where $\phi(\theta;\alpha)$ is the stripe wave phase defined by $\cos \phi(\theta;\alpha) \equiv -\cot(\theta) \cot(\alpha)$.

Thus, the total fraction of magnetar power which remains in the ordered magnetic field following reconnection at the termination shock is
\begin{equation}
f_{\rm j}(\alpha) 
= \frac{\int
\left({dL_{\rm e}}/{d\Omega}\right) 
\chi
d\Omega}{\int
\left({dL_{\rm e}}/{d\Omega}\right) d\Omega}
= \frac{3}{2} \int_0^{\frac{\pi}{2}} \chi(\theta; \alpha) \sin^3 \theta d\theta .
\label{eq:fj}
\end{equation}
Similarly, the thermalized energy fraction is $f_{\rm th}(\alpha) = 1-f_{\rm j}(\alpha)$.
We find that $f_{\rm th}$ is well approximated (to within an accuracy of a couple percent) by 
\begin{equation}
f_{\rm th}(\alpha) \approx \frac{ \left[ 1 + \left({\pi}/{2}\right)^{-4} b \right]^{1/4} \alpha}{ \left( b + \alpha^4 \right)^{1/4}}
\simeq \frac{1.025 \alpha}{\left( 0.636 + \alpha^4 \right)^{1/4}} ,
\label{eq:fth_fit}
\end{equation}
where $\alpha$ is in radians and in the second equation $b \simeq 0.636$.
The model thus implies that any oblique rotator will partition its spin-down power into both an ordered magnetic and a thermal component, with thermalization increasing for greater misalignment angles $\alpha$.

\begin{figure}
\centering
\includegraphics[width=0.5\textwidth]{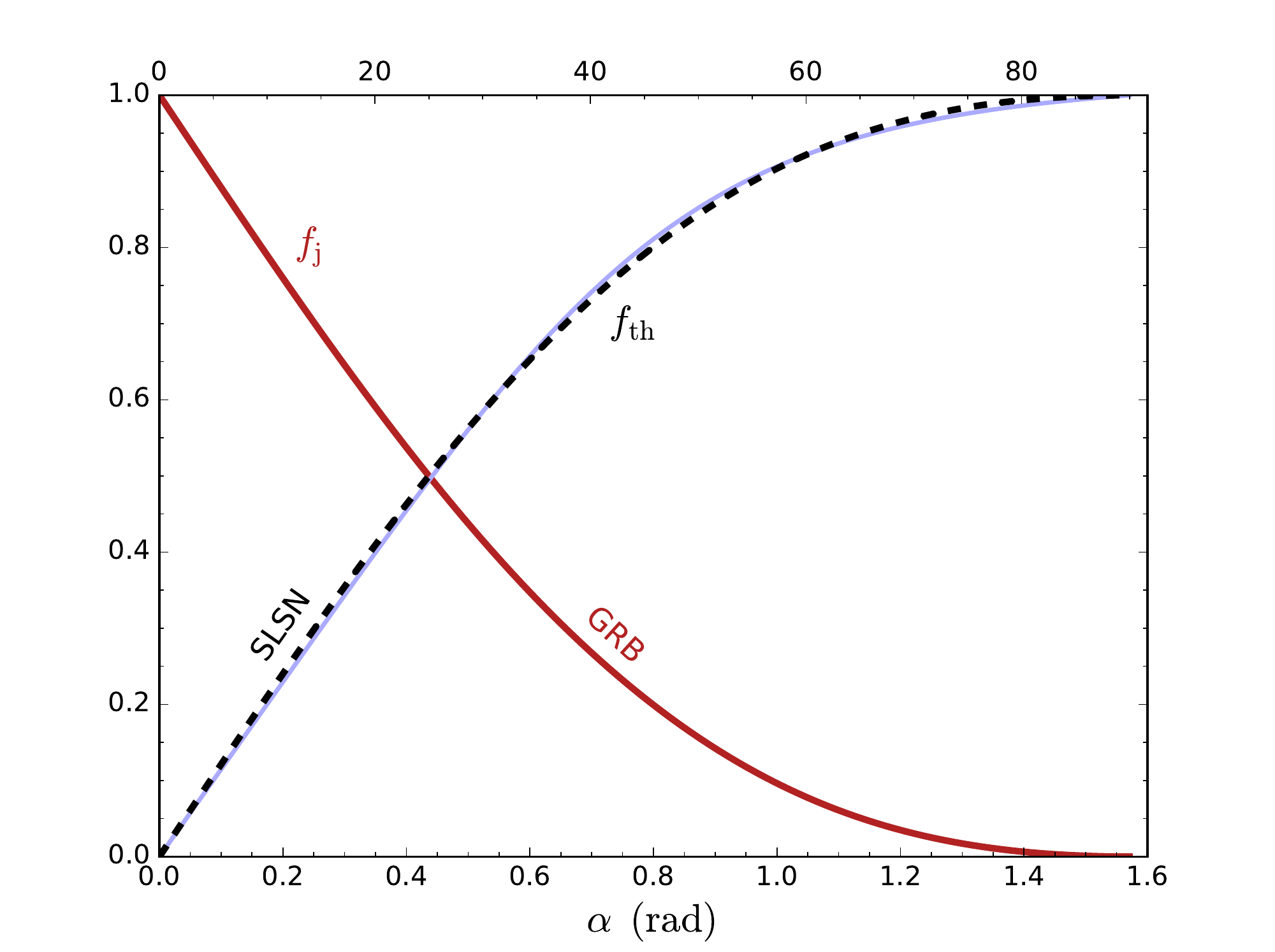}
\caption{Fraction of the spin-down luminosity of the magnetar available for powering an ordered, magnetically-dominated jet $f_{\rm j}$ (solid red; equation~\ref{eq:fj}) versus the complementary fraction $f_{\rm th}=1-f_{\rm j}$ (dashed black) which is thermalized due to forced reconnection in the striped wind, shown as a function of the misalignment angle between rotation and magnetic axes, $\alpha$. 
$f_{\rm th}$ is well approximated by 
equation~\ref{eq:fth_fit} (solid blue).
In this model, a mis-aligned magnetar can simultaneously power both a luminuous SN and a jetted GRB.} \label{fig:fmfth}
\end{figure}

It is therefore natural to interpret $f_{\rm j}$ --- the fraction of the energy remaining in an ordered toroidal magnetic field --- as that which may contribute to a collimated jet component (GRB), while $f_{\rm th}$ is the complementary power energizing the SN ejecta which may contribute to powering to the isotropic thermal emission (SN).
The mis-alignment angle can thus be observationally inferred by identifying the thermalization fraction with $f_{\rm th} \approx E_{\rm SLSN} / (E_{\rm SLSN} + E_{\rm GRB} )$ and inverting equation~(\ref{eq:fth_fit}), yielding
\begin{equation} \label{eq:alpha_fit}
\alpha \approx {0.893 f_{\rm th}} {\left( 1.105 - f_{\rm th}^4 \right)^{-1/4}} \, {\rm rad} .
\end{equation}

The simplified picture outlined above assumes that magnetic energy is only dissipated through reconnection of a striped wind at the termination shock.  However, other forms of dissipation related to MHD instabilities (e.g.~kink or sausage), may operate on larger scales throughout the nebula as well (e.g.~\citealt{Begelman98,Porth&Komissarov13,Zrake&Arons17}), depending in part on how effectively the build-up of toroidal flux is ``relieved" by the escape of a successful polar jet.  The details of such a thermalization processes are more complex, and we briefly discuss their affect on the jet component in \S\ref{subsec:JetSignature}.

Finally, note that the jet model we develop in the following section can equally be applied to black-hole (accretion-powered) engines. In this case, we expect $\ell \approx 5/3$ in equation~(\ref{eq:Le}), as set by the rate of mass fall-back of marginally-bound stellar debris following the SN\footnote{However, see \citet{Tchekhovskoy&Giannios15}, who argue that the jet power may be set by the rate of accumulation of magnetic flux onto the black hole, rather than the accretion rate, in which case the time-dependence of the engine luminosity will be more complicated.}. 
The engine energy is related to the total fall-back mass $M_{\rm fb}$ according to $E_{\rm e} = \epsilon_{\rm fb} M_{\rm fb} c^2$, where $\epsilon_{\rm fb}< 1$ is an efficiency factor for producing a relativistic jet or disk wind. 
The engine timescale $t_{\rm e}$ is generally set by the gravitational free-fall timescale of the progenitor star, $t_{\rm e} \sim 1/\sqrt{ G \bar{\rho} }$, where $\bar{\rho}(r)$ is the average density of the enclosed mass within radius $r$ of the stellar progenitor \citep[e.g.][]{Dexter&Kasen13}.  This timescale is substantially shorter in the case of the compact Wolf-Rayet progenitors responsible for SNe Ic-bl, compared to the more radially-extended outer layers of blue or red supergiants.
As in the magnetar case, accretion-powered engines may exhibit a dichotomy between a bipolar relativistic outflow and slower wide-angle wind, which contribute to powering the GRB and isotropic SN, respectively.  However, the energetic partitioning between these two components depends in a more complex manner on the details of the accretion model, such as the angular momentum distribution (\citealt{Dexter&Kasen13}) and magnetic flux (\citealt{Tchekhovskoy&Giannios15}) of the fall-back mass.

\section{Weak Jet Break-out} \label{sec:Jet}
Jet propagation and break-out from stationary stellar progenitors have been extensively studied in the context of GRB \citep{Aloy+00,MacFadyen+01,Zhang+03,Morsony+07,LopezCamara+13,Bromberg&Tchekhovskoy16}. 
The emerging picture is of a self-collimated relativistic outflow confined by an isobaric cocoon of hot shocked matter which spills around the jet-head \citep[e.g.][]{Bromberg+11}.

Central-engine powered SNe generally require long-lived engine activity ($\sim 10$ d) compared to those of typical LGRB ($\sim 100$ s). For a fixed engine energy budget, this implies dramatically lower power.
Such low-luminosity jets may therefore not manage to burrow their way out of the surrounding stellar progenitor faster than the SN shock-front, resulting in dramatically different ambient conditions for jet propagation.
In the following, we extend analytic jet models to low-luminosity jets propagating within an exploding stellar profile and expanding SN ejecta. 
The latter has been previously considered by \cite{Quataert&Kasen12}, but has not received attention beyond that work.
Our main finding is that weak jets --- of the kind expected to accompany SLSNe if the central engine partitions comparable energy into jetted and thermal components (see \S\ref{sec:MagnetarModel}) --- may successfully break out of the expanding SN ejecta at late times (cf. equations~\ref{eq:Ee_min},\ref{eq:tbo}).

Although the magnetar picture presented in \S\ref{sec:MagnetarModel} dictates a magnetically dominated jet, recent numerical simulations by \cite{Bromberg&Tchekhovskoy16} illustrate that such jets behave similarly to hydrodynamic jet models during their escape from the star.  We therefore follow \cite{Bromberg+11} and adopt a hydrodynamic collimated-jet model. Our results differ from those of \cite{Quataert&Kasen12} primarily due to the fact that these authors assumed an uncollimated jet model.
Readers uninterested in the model details are encouraged to jump to \S\ref{subsec:BreakoutConditions} (in particular \ref{subsubsec:breakout_from_ejecta}), where we present  our main findings.

\subsection{Jet Propagation Model}
\subsubsection{Ambient Density Profile}
\label{subsec:ambientdensity}
We assume an initial stellar progenitor of mass $M_\star$, radius $R_\star$, described by a power-law density profile $\rho \propto r^{-w}$ with $w<3$, such that
\begin{equation} \label{eq:rho_stellar}
\rho(r,t=0) = \frac{(3-w)M_\star}{4 \pi R_\star^3} \left(\frac{r}{R_\star}\right)^{-w} .
\end{equation}
A SN explosion of energy $E_{\rm sn} \sim 10^{51} \, {\rm erg}$ generates an outgoing shock front which propagates as 
\begin{equation}
R_{\rm sh}(t \leq t_\star) = R_\star \left( \frac{t}{t_\star} \right)^{2/(5-w)},
\end{equation}
where 
\begin{align}
t_{\star} &= \sqrt{\frac{2M_\star R_{\star}^2}{(5-w)^2 E_{\rm sn}}}
\nonumber \\
&\simeq 173 \, {\rm s} \, 
\left(\frac{R_\star}{10^{11} \, {\rm cm}}\right) \left(\frac{M_\star}{5 M_\odot}\right)^{1/2} \left(\frac{E_{\rm sn}}{10^{51} \, {\rm erg}}\right)^{-1/2}
\label{eq:tstar}
\end{align}
is the time at which the shock front reaches the stellar surface. Up to numerical factors of order unity (e.g. \citealt{Chevalier76}) this is the standard Sedov-Taylor solution.

We schematically follow \cite{Chevalier&Soker89} who focus on the case of a progenitor density power-law of $w = 17/7 \simeq 2.43$. This profile is particularly relevant as it is characteristic of stripped envelope stars which are the possible progenitors of LGRB, Type Ib/c SNe, and Type I SLSNe, and are well modeled by $\rho \sim r^{-2.5}$ \citep{Woosley&Heger06}.
The precise value of $w=17/7$ is convenient as it considerably simplifies the solution by reducing to the so-called Primakoff blast-wave.
In this case, the density increases linearly up to the radius of the outwardly-propagating shock front, and can be expressed as
\begin{align} 
\rho(r<R_{\rm sh}, t < t_{\star})
&= 7 \rho(R_{\rm sh}, t = 0) \left(\frac{r}{R_{\rm sh}}\right) 
\nonumber \\
&= \frac{M_\star}{\pi R_\star^3} \left(\frac{r}{R_\star}\right) \left(\frac{t}{t_\star}\right)^{-8/3} .
\label{eq:rho_blastwave}
\end{align}
The unshocked region ahead of the blast-wave ($r> R_{\rm sh}$) remains unperturbed and is still described by the initial stellar density profile (equation~\ref{eq:rho_stellar}).

After reaching the progenitor's surface at $t=t_\star$, the shock is accelerated in the dilute stellar atmosphere, the details of which we ignore here. Eventually, at late times ($t \gg t_\star $) the SN ejecta expands homologously, such that $v=r/t$ and
\begin{equation} \label{eq:rho_ejecta}
\rho(r \leq v_{\rm ej}t, t \gg t_\star ) = \zeta_\rho \frac{M_{\rm ej}}{(v_{\rm ej} t)^3} \left(\frac{r}{v_{\rm ej} t}\right)^{-\delta} ,
\end{equation}
with $\delta = 1$ a typical value \citep{Chevalier&Soker89}.
The characteristic ejecta velocity is given by
\begin{equation} \label{eq:vej}
v_{\rm ej} = \zeta_v \sqrt{E_{\rm sn}/M_{\rm ej}} ,
\end{equation}
and the numerical constants $\zeta_\rho$, $\zeta_v$ are defined as
\begin{equation} \label{eq:zeta_rho_v}
\zeta_\rho = \frac{3-\delta}{4\pi}, ~~~
\zeta_v = \sqrt{\frac{2(5-\delta)}{3-\delta}} .
\end{equation}

\begin{figure}
\includegraphics[width=0.5\textwidth]{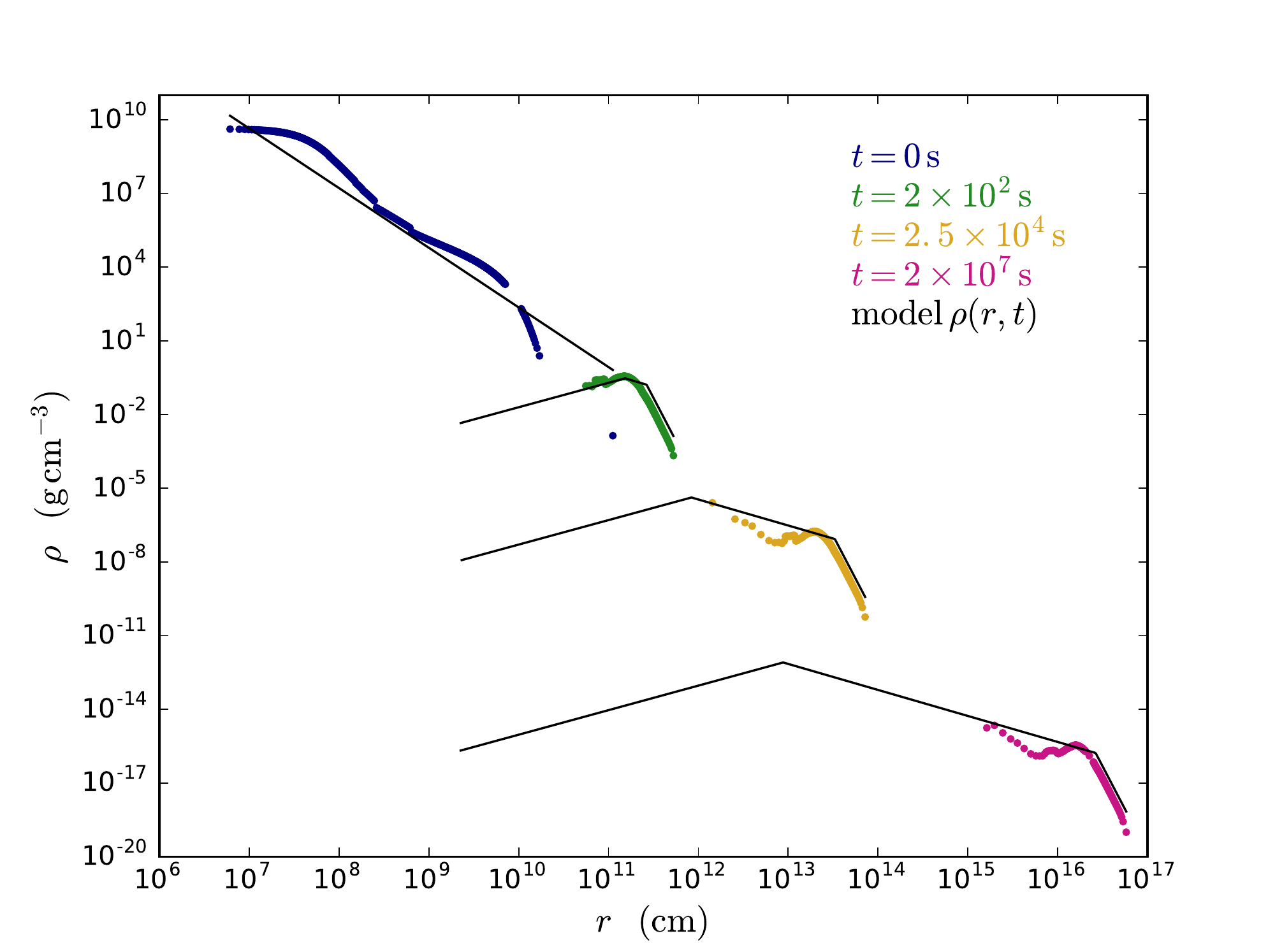}
\caption{Density profile snapshots at different times (colored points) of a $10 M_\odot$ carbon-oxygen core numerically exploded (with $10^{52} \, {\rm erg}$) in a 1D hydrodynamics code. Our analytic model (black lines; \S\ref{subsec:ambientdensity}) readily reproduces the large-scale features of the pre-SN progenitor and subsequent expanding SN ejecta.
}  \label{fig:density}
\end{figure}

We compare our analytic model for $\rho(r,t)$ with 1D numerical simulation results in Fig.~\ref{fig:density}. 
The numerical profiles were calculated using the 1D implicit hydrodynamic code {\tt KEPLER} \citep{Weaver78}. Starting from a presupernova model of a $M_\star = 10 M_\odot$ carbon-oxygen core from the same code \citep{Sukhbold&Woosley14} the explosion of $E_{\rm sn} = 10^{52} \, {\rm erg}$ is calculated using the moving inner boundary method (i.e. `piston' method, \citealt{Woosley&Weaver95}).
Our analytic model is fully described by specifying $M_\star$, $R_\star$ and $E_{\rm sn}$, which we take directly from the simulation setup.
While our simplified analytic prescription cannot reproduce many of the finer features present in the numerical results, the large-scale structure of the density profiles is reasonably approximated by our model.

\subsubsection{Jet Propagation}
The jet dynamics are determined by the dimensionless jet luminosity  parameter
$\tilde{L} \equiv {L_{\rm j}}/{A_{\rm j} \rho c^3} $,
where $A_{\rm j}$ is the jet head cross-section as it burrows through the star, $L_{\rm j} = f_{\rm j} L_{\rm e}/2$ is the {\it one-sided} jet luminosity, and $L_{\rm e}$ is the total engine luminosity (equation~\ref{eq:Le}). 
A jet which obeys $\tilde{L} \lesssim \gamma_{\rm j}^{4/3}$, where $\gamma_{\rm j}$ is the jet Lorentz factor, remains self-collimated as it propagates through the star. Additionally, a jet with $\tilde{L} \ll 1$ moves through the star at a non-relativistic velocity.  In these limits, the dimensionless jet power --- as a function of the jet-head position within the ambient medium $z_{\rm h}$ --- is given by \citep{Bromberg+11}
\begin{equation} \label{eq:Ltilde}
\tilde{L}(z_{\rm h}) = \zeta_{\tilde{L}}^{2/3} \left(\frac{L_{\rm j} \gamma_{\rm j}^4}{\rho\left(z_{\rm h}\right) z_{\rm h}^2 c^3}\right)^{2/3}
\equiv \tilde{L}_\star \zeta_{\tilde{L}}^{2/3} \left(\frac{\rho\left(z_{\rm h}\right) z_{\rm h}^2 c^3}{M_\star R_\star^2 t_\star^{-3}}\right)^{-2/3} .
\end{equation}
Here $\zeta_{\tilde{L}}$ is an order unity numerical factor derived in Appendix~\ref{sec:Appendix_Jet} (equation~\ref{eq:zeta_Ltilde}),
and in the second equation we have defined 
\begin{align} \label{eq:Ltilde_star}
\tilde{L}_\star \equiv \left( \frac{L_{\rm j} \gamma_{\rm j}^4}{M_\star R_\star^2 t_\star^{-3}}\right)^{2/3}
&\simeq 0.89 \times \left(\frac{L_{\rm j}}{10^{48} \, {\rm erg \, s}^{-1}}\right)^{2/3} 
\left(\frac{\gamma_{\rm j}}{2}\right)^{8/3}
\\ \nonumber
&\times \left(\frac{M_\star}{5 M_\odot}\right)^{1/3}
\left(\frac{R_\star}{10^{11} \, {\rm cm}}\right)^{2/3}
\left(\frac{E_{\rm sn}}{10^{51} \, {\rm erg}}\right)^{-1}
\end{align}
as a convenient measure of the jet luminosity which depends only on $L_{\rm j}$ and the stellar parameters (and not on $z_{\rm h}$). In \S\ref{subsec:BreakoutConditions} we show that $\tilde{L}_\star$ differentiates between various jet break-out regimes.

The jet-head velocity $v_{\rm h}$ in the self-collimated non-relativistic limit is related to the dimensionless luminosity by \citep[e.g.][]{Matzner03}
\begin{equation} \label{eq:vh}
v_{\rm h} = \frac{c}{1 + \tilde{L}^{-1/2}} \approx c \left[\tilde{L}(z_{\rm h})\right]^{1/2} .
\end{equation}
This is a differential equation for the vertical location of the jet-head $z_{\rm h}(t)$, which is readily solved for any given ambient density and jet luminosity. For a power-law density profile the solution is derived in Appendix~\ref{sec:Appendix_Jet}, and acquires the simple form $z_{\rm h} = \zeta_z v_{\rm h} t$.


Though subject to many uncertainties, the jet Lorentz factor appearing in equation~(\ref{eq:Ltilde}) can be constrained from inferred GRB jet-opening angles, since $\gamma_{\rm j} \sim \theta_{\rm j}^{-1}$ \citep{Bromberg+11}. \cite{Mizuta&Ioka13} numerically calibrate this relation to $\gamma_{\rm j} \approx 1/5\theta_{\rm j}$, and infer typical GRB Lorentz factors of $\gamma_{\rm j} \approx 2$-$3$. We therefore adopt $\gamma_{\rm j} = 2$ as a fiducial value throughout this work.

\cite{Bromberg&Tchekhovskoy16} show that magnetically dominated jets are susceptible to the global kink-instability when the dimensionless parameter $\Lambda \lesssim 1$, where
\begin{equation}
\Lambda \equiv 10 \eta \frac{t_{\rm kink}}{t_{\rm dyn}} = 10 \eta \frac{2\pi \gamma_{\rm j} R_{\rm j} / v_{\rm A}}{z_{\rm h} / (c-v_{\rm h})} ,
\label{eq:kink}
\end{equation}
$R_{\rm j} \approx \sqrt{A_{\rm j}/\pi}$ is the cylindrical radius of the jet head, $v_{\rm A} \approx c$ is the Alfv{\'e}n speed, and $\eta$ is an order unity ignorance factor which \cite{Bromberg&Tchekhovskoy16} numerically find satisfies $0.5 \lesssim \eta \lesssim 1$.
Expressing the jet-head velocity and cross-section as a function of $\tilde{L}$ using the general unapproximated expressions (i.e. relaxing the assumption $\tilde{L} \ll 1$), we find after some algebraic manipulation that
\begin{equation} \label{eq:Lambda}
\Lambda = 20 \sqrt{\pi} \eta \gamma_{\rm j}^{-1} \zeta_{\tilde{L}}^{-1/2} \tilde{L}^{1/4}
\end{equation}

\subsection{Jet Breakout Conditions} \label{subsec:BreakoutConditions}
\subsubsection{from Stellar Surface ($\tilde{L}_\star \gtrsim 1$)}
For sufficiently large luminosities, the jet burrows out of the stellar progenitor before the SN blast-wave has any significant effect on the outer layers of the star (i.e. $t_{\rm bo} \ll t_\star$). This is the usual scenario considered in the literature \citep[e.g.][]{Bromberg+11}, where the jet can be treated as propagating within a hydrostatic stellar environment described by equation (\ref{eq:rho_stellar}). Equating $R_\star$ with $\zeta_z v_{\rm h}(t_{\rm bo}) t_{\rm bo}$ yields an estimate of the jet breakout time in this limit,
\begin{align} \label{eq:tbo_stellar}
t_{\rm bo} &= 
\zeta_\rho^{1/3} \zeta_z^{-1} \zeta_{\tilde{L}}^{-1/3} \tilde{L}_\star^{-1/2} t_\star
\simeq 0.19 \tilde{L}_\star^{-1/2} t_\star
\\ \nonumber
&\simeq 7.6 \, {\rm s} \, \left(\frac{L_{\rm j}}{10^{50} \, {\rm erg \, s}^{-1}}\right)^{-1/3} \left(\frac{\gamma_{\rm j}}{2}\right)^{-4/3} \left(\frac{M_\star}{5 M_\odot}\right)^{1/3} \left(\frac{R_\star}{10^{11} \, {\rm cm}}\right)^{2/3}
\end{align}
where $\zeta_\rho$, $\zeta_z$, $\zeta_{\tilde{L}}$ are calculated using equations~(\ref{eq:zeta_rho_v},\ref{eq:zeta_Ltilde},\ref{eq:zeta_z}) with the understanding that ``$\delta$"$=w$ in the considered case (density described by equation~\ref{eq:rho_stellar}). 
Equation (\ref{eq:tbo_stellar}) is therefore applicable in the regime $\tilde{L}_\star \gg 1$, characteristic of the jet luminosities of normal long duration GRB (equation~\ref{eq:Ltilde_star}).

The jet fails to break out in this regime if the break-out time $t_{\rm bo}$ exceeds the engine duration $t_{\rm e}$. Combining equations (\ref{eq:tbo_stellar}) and (\ref{eq:Le}), successful breakout requires a minimum engine duration of
\begin{align} \label{eq:te_min}
t_{\rm e} \gtrsim 
2.9 \, {\rm s} \,  \left(\frac{f_{\rm j} E_{\rm e}}{10^{52} \, {\rm erg}}\right)^{-1/2} \left(\frac{\gamma_{\rm j}}{2}\right)^{-2} \left(\frac{M_\star}{5 M_\odot}\right)^{1/2}\left(\frac{R_\star}{10^{11} \, {\rm cm}}\right) .
\end{align}

When $\tilde{L}_\star \sim 1$ the jet escape timescale is comparable to the time required for the SN shock to reach the stellar surface, in which case $t_{\rm bo}$ differs from equation (\ref{eq:tbo_stellar}). In the marginal case defining the boundary between jet-breakout from the original stellar surface $R_\star$ versus from an expanding SN envelope, i.e. $t_{\rm bo} = t_\star$, the jet can be considered as propagating entirely within the density profile (\ref{eq:rho_blastwave}). This marginal case occurs for a critical jet luminosity\footnote{Note that the numerical values of $\zeta_\rho$, $\zeta_{\tilde{L}}$ and $\zeta_z$ differ slightly between equations (\ref{eq:tbo_stellar}) and (\ref{eq:Lstar_bo_tstar}). In the latter case, these constants are calculated from equations~(\ref{eq:zeta_rho_v},\ref{eq:zeta_Ltilde},\ref{eq:zeta_z}) with the understanding that $\delta=-1, \beta=8/3$, as implied by equation~(\ref{eq:rho_blastwave}).}
\begin{equation} \label{eq:Lstar_bo_tstar}
\tilde{L}_\star(t_{\rm bo}=t_\star) = \zeta_\rho^{2/3} \zeta_z^{-2} \zeta_{\tilde{L}}^{-2/3} 
\simeq 0.37 ,
\end{equation}
or, by virtue of equation~(\ref{eq:Ltilde_star}), $L_{\rm j}(t_{\rm bo}=t_\star) \simeq 3 \times 10^{47} \, {\rm erg \, s}^{-1}$.

Finally, in the weak jet regime ($\tilde{L}_\star \ll 1$) relevant to long-lived engines capable of powering SLSNe, the jet propagates predominantly within the expanding SN ejecta described by equation (\ref{eq:rho_ejecta}). We focus on this regime for the remainder of this section.

\subsubsection{from Expanding Ejecta ($\tilde{L}_\star \ll 1$)} \label{subsubsec:breakout_from_ejecta}
We consider two criterion for the jet to successfully escape from the expanding SN ejecta.  First, we require that the jet head overtakes the expanding ejecta $z_{\rm h}(t_{\rm bo}) \gtrsim v_{\rm ej}t_{\rm bo}$ at breakout, which is essentially equivalent to the requirement that the head velocity exceed the ejecta velocity $v_{\rm h} \gtrsim v_{\rm ej}$ \citep{Quataert&Kasen12}.
Second, the jet must also satisfy the kink-stability criterion $\Lambda \gtrsim 1$ (equation~\ref{eq:kink}).  Combining these criterion, successful jet break-out occurs if both of the following conditions on its luminosity are satisfied,
\begin{align}
&v_{\rm h} \gtrsim \frac{v_{\rm ej}}{\zeta_z}
&\Rightarrow& 
\tilde{L} \gtrsim \tilde{L}_{\rm crit}  = \left(\frac{v_{\rm ej}}{\zeta_zc}\right)^2 
\sim 10^{-3} \left(\frac{v_{\rm ej}}{10^9\,{\rm cm \, s}^{-1}}\right)^2
\label{eq:vh_bo_criterion} \\
&\Lambda \gtrsim 1
&\Rightarrow& 
\tilde{L} \gtrsim \tilde{L}_{\rm crit}  = \left(\frac{\gamma_{\rm j} \zeta_{\tilde{L}}^{1/2}}{20 \sqrt{\pi} \eta}\right)^4 
\sim 10^{-4} \left(\frac{\gamma_{\rm j}}{2}\right)^4 \eta^{-4} \label{eq:kinkinstability_criterion}
,
\end{align}
or, in terms of the physical jet luminosity (equation \ref{eq:Ltilde}),
\begin{equation} \label{eq:Lj_min}
L_{\rm j}(t) \gtrsim 
\zeta_\rho \zeta_{\tilde{L}}^{-1} \gamma_{\rm j}^{-4} c^3 \tilde{L}_{\rm crit}^{3/2}  \frac{M_{\rm ej}}{v_{\rm ej} t} ,
\end{equation}
where we have substituted equation~(\ref{eq:rho_ejecta}) for the ejecta density, and evaluated $\rho z^2$ at the ejecta front ($z = v_{\rm ej} t$) where it attains its maximal value, and the breakout condition is hardest to satisfy.

For typical parameters, the more stringent break-out threshold $\tilde{L}_{\rm crit}$ is set by the dynamical condition (\ref{eq:vh_bo_criterion}) on the head velocity instead of kink instability.  Also note that our assumption at the beginning of this section that the jet propagates through the star at non-relativistic velocities in the collimated regime is also justified because  $\tilde{L}_{\rm crit} \ll 1, \gamma_{\rm j}^{4/3}$ are satisfied for all reasonable parameters.

In both the magnetar and fall-back scenarios the central engine power (equation~\ref{eq:Le}) decays at times $t \gg t_{\rm e}$ as $L_{\rm j} \sim t^{-\ell}$ with $\ell > 1$.  In such cases the LHS of equation~(\ref{eq:Lj_min}) decreases faster with time than the RHS, indicating that successful jet break-out {\it can only occur prior to the engine shut-off time $t_{\rm e}$}.  More precisely, condition~(\ref{eq:Lj_min}) must be satisfied before the critical time $t_{\rm e} / (\ell-1)$ at which $\partial \ln L_{\rm j} / \partial \ln t = -1$.  By substituting equation~(\ref{eq:Le}) at this critical time in equation~(\ref{eq:Lj_min}) and expressing the jet energy in terms of the total engine energy, $E_{\rm j} = f_{\rm j} E_{\rm e} / 2$ (the factor of two accounts for the bipolar nature of the jet), this breakout condition may be recast entirely as a constraint on the total engine energy $E_{\rm e}$ as follows
\begin{align} \label{eq:Ee_min}
E_{\rm e} \gtrsim E_{\rm e,min} &= 2 f_{\rm j}^{-1} f_\ell \zeta_\rho \zeta_{\tilde{L}}^{-1} \gamma_{\rm j}^{-4} \tilde{L}_{\rm crit}^{3/2}  \frac{M_{\rm ej} c^3}{v_{\rm ej}}
\nonumber \\
&\simeq 0.195 E_{\rm sn} \left(\frac{\gamma_{\rm j}}{2}\right)^{-4} f_{\rm j}^{-1},
\end{align}
where 
$f_\ell \equiv \ell^\ell (\ell-1)^{(1-\ell)} \simeq 4$ and in the second line we have used the threshold critical luminosity $\tilde{L}_{\rm crit}$ from equation~(\ref{eq:vh_bo_criterion}) along with equation~(\ref{eq:vej}).  
In magnetar models, this lower limit on the engine energy implies a maximum birth spin-period for successful jet breakout (equation~\ref{eq:Ee_magnetar}) of 
\begin{equation} \label{eq:P0_max}
P_0 \lesssim 11.3 \, {\rm ms} \left(\frac{E_{\rm sn}}{10^{51}\,{\rm erg}}\right)^{-1/2} \left(\frac{\gamma_{\rm j}}{2}\right)^{2} f_{\rm j}^{1/2}.
\end{equation}  

For engine energies $E_{\rm e} > E_{\rm e,min}$ exceeding this minimum value, equation~(\ref{eq:Lj_min}) yields a
break-out time of
\begin{align} \label{eq:tbo}
t_{\rm bo} &= \zeta_{\rho} \zeta_{\tilde{L}}^{-1} \zeta_{z}^{-3} \zeta_{v}^{2} \gamma_{\rm j}^{-4} E_{\rm sn} / L_{\rm j}
\nonumber \\
&\simeq 2.4 \times 10^3 \, {\rm s} \, 
\left(\frac{\gamma_{\rm j}}{2}\right)^{-4} \left(\frac{E_{\rm sn}}{10^{51} \, {\rm erg}}\right) \left(\frac{L_{\rm j}}{10^{46} \, {\rm erg \, s}^{-1}}\right)^{-1} .
\end{align}
Equation~(\ref{eq:tbo}) is identical in form to the timescale for shock break-out from spherical pulsar wind nebulae derived by \cite{Chevalier&Fransson92} \citep[see also][]{Chevalier05}, differing only by numerical constants $\propto \gamma_{\rm j}^{-4}$.

\section{Off-axis Jet Signature} \label{subsec:JetSignature}
The results of \S\ref{sec:Jet} illustrate that jets with relatively low luminosities, of the same order of magnitude as the engine luminosities needed to power SLSNe, may successfully escape their SN ejecta on timescales comparable or less than the engine lifetime (equation~\ref{eq:tbo}).  This is consistent with the discovery of GRB 111209A in association with the luminous SN 2011kl \citep{Greiner+15}.  

In this section, we turn to the next natural question --- what are the observable signatures of such jets?   We focus on the most common case in which the jet axis is not aligned with the observer's line of sight, precluding an associated GRB.
In \S\ref{sec:cocoon} we explore cocoon break-out emission, which can give rise to short-duration ($\sim$hr long) UV transients. We then propose a novel model whereby radiation from a thermally-driven wind launched off the jet-ejecta interface can produce an early ($\sim$several day) maxima in SLSNe light-curves (\S\ref{subsubsec:jet_energized_emission}). Finally, in \S\ref{sec:RadioAfterglow}, we examine radio afterglow constraints on observed SLSNe.

\subsection{Cocoon Breakout Emission}
\label{sec:cocoon}
We consider first the signature of the hot cocoon which encases the jet after it breaks out of the stellar surface.  
As discussed by \citet{Nakar&Piran16}, the most promising cocoon signature is that arising from the isotropic non-relativistic {\it shocked-star} component of the cocoon (as opposed to the dilute, relativistic shocked-jet component). This component expands with an average bulk velocity of $v_{\rm c} \sim \sqrt{P_{\rm c}/\bar{\rho}} \approx \gamma_{\rm j}^{-1} v_{\rm h}$, and in the stationary-stellar model produces emission from the cooling envelope \citep{Nakar&Piran16}.
By contrast, in the case of breakout from an expanding SN-envelope considered here, the jet overtakes the ejecta only once the head velocity reaches $v_{\rm h} \sim v_{\rm ej}$ (condition \ref{eq:vh_bo_criterion}).  This implies that the shocked-star component of the cocoon (which may be more appropriately named shocked SN-ejecta in this case) will be expanding at velocities $< v_{\rm ej}$ and therefore cannot itself break-out ahead of the expanding ejecta.  For this reason, we do not expect a signature from the bulk shocked-star component of the cocoon in the late-breakout scenario, unless potentially in rare cases when the jet axis aligns with the observer's line of site.

An alternative isotropic signature may arise if some fraction of the {\it shocked-jet} cocoon component attains trans-relativistic (as opposed to exclusively ultra-relativistic) velocities, thereby managing to overtake the surrounding SN-ejecta while also producing emission that is not highly relativistically beamed.  Following the notation of \cite{Nakar&Piran16}, we assume that a fraction $f_{\Gamma\beta,1} \sim 0.1$ of the cocoon's energy is deposited into a trans-relativistic component expanding at proper-velocities $\Gamma\beta \sim 1$, as appropriate for the case of significant, but not complete mixing between the shocked-star and shocked-jet. The main free parameter of the model is the expansion velocity, $\beta_{\rm c,j}c$, the value of which we take fiducially to be $\beta_{\rm c,j} = 0.7$.

The cocoon's thermal energy at breakout, 
$E_{\rm c} \approx 2 L_{\rm j} t_{\rm bo}$,
implies a trans-relativistic cocoon component mass 
\begin{align}
M_{\rm c,j} \approx 2 f_{\Gamma\beta,1} L_{\rm j} t_{\rm bo} / \left({\beta_{\rm c,j} c}\right)^{2} ,
\end{align}
of order $\sim 10^{-5} M_\odot$ for typical parameters.

We now consider two cases, depending on the shock break-out time $t_{\rm bo}$.  For relatively short break-out times $t_{\rm bo} \ll 1$~d, the optical depth of the cocoon after one initial expansion timescale (over which the cocoon wraps spherically around the SN ejecta) exceeds the value $\sim \beta_{\rm c,j}^{-1}$.  In this case the initial photon diffusion time exceeds the expansion time, and so the bulk of the radiation is initially trapped and the observed luminosity will peak only after further radial expansion, at times $t \gg  t_{\rm bo}$.  This situation is  well described by the standard cooling envelope emission model \citep[e.g.][]{Nakar&Piran16}, resulting in emission for a characteristic duration of
\begin{align}
t_{\rm pk} = \sqrt{\frac{3 \kappa M_{\rm c,j}}{4\pi c^2 \beta_{\rm c,j}}} 
\simeq 9.2 \times 10^2 \, {\rm s} \, 
&\left(\frac{E_{\rm sn}}{10^{51}\,{\rm erg}}\right)^{1/2} \left(\frac{\gamma_{\rm j}}{2}\right)^{-2}
\\ \nonumber &\times 
 \left(\frac{f_{\Gamma\beta,1}}{0.1}\right)^{1/2} \left(\frac{\beta_{\rm c,j}}{0.7}\right)^{-3/2} 
,
\end{align}
with a peak (bolometric) luminosity of 
\begin{align}
L_{\rm pk} &\approx \frac{E_0 t_{\rm exp,0}}{t_{\rm pk}^2}
\simeq 6.5 \times 10^{43} \, {\rm erg \, s}^{-1} \, 
\left(\frac{L_{\rm j}}{10^{46} \, {\rm erg \, s}^{-1}}\right)^{-1}
\\ \nonumber &\times 
\left(\frac{E_{\rm sn}}{10^{51} \, {\rm erg}}\right) \left(\frac{\gamma_{\rm j}}{2}\right)^{-14/3}
\left(\frac{v_{\rm ej}}{10^9 \, {\rm cm \, s}^{-1}}\right)  \left(\frac{f_{\Gamma\beta,1}}{0.1}\right)^{1/3} \left(\frac{\beta_{\rm c,j}}{0.7}\right)^2 .
\end{align}
Here $E_0 = (V_{\rm c,j} / V_0 )^{1/3} E_{\rm c,j} \approx (f_{\Gamma\beta} \gamma_{\rm j}^{-2} / 28)^{1/3} E_{\rm c,j}$ is the thermal energy of the cocoon following one expansion time $t_{\rm exp,0} = v_{\rm ej} t_{\rm bo}/\beta_{\rm c,j} c$ after breakout, and $V_{\rm c,j} = f_{\Gamma\beta,1} V_{\rm c}$ is the volume required by pressure equilibrium inside the cocoon (which also implies that $\theta_{\rm c,j} = \theta_{\rm c} f_{\Gamma\beta,1}^{1/2} \approx \gamma_{\rm j}^{-1} f_{\Gamma\beta,1}^{1/2}$).

In the opposite limit of long breakout times ($t_{\rm bo} \gtrsim 1$ d), the shocked ejecta becomes transparent to thermal radiation prior to significant radial expansion.  The standard cooling envelope model does not apply because the thermal energy is radiated before the shocked ejecta has time to expand quasi-spherically around the SN ejecta.  In particular, for jet luminosities in the range
\begin{align} \label{eq:L_cj_heirarchy_condition}
L_{\rm min} < L_{\rm j} &< L_{\rm min} \times \theta_{\rm c,j}^{-1} (1-\beta_{\rm c,j})^{-1/2} 
\\ \nonumber
&\simeq 11.6 L_{\rm min} \left(\frac{\gamma_{\rm j}}{2}\right) \left(\frac{f_{\Gamma\beta,1}}{0.1}\right)^{-1/2} \left(\frac{1-\beta_{\rm c,j}}{0.3}\right)^{-1/2}
\end{align}
where
\begin{align}
L_{\rm min} \simeq 
6.7 &\times 10^{43} \, {\rm erg \, s}^{-1}
\left(\frac{v_{\rm ej}}{10^9 \, {\rm cm \, s}^{-1}}\right) \left(\frac{E_{\rm sn}}{10^{51} \, {\rm erg}}\right)^{1/2}  
\\ \nonumber
&\times \left(\frac{\gamma_{\rm j}}{2}\right)^3 \left(\frac{\kappa}{0.2 \, {\rm cm}^2 \, {\rm g}^{-1}}\right) \left(\frac{\beta_{\rm c,j}(1-\beta_{\rm c,j})}{0.21}\right)^{1/2}
\end{align}
the expansion timescale at breakout --- while longer than the timescale for radiative diffusion in the azimuthal direction --- is still shorter than the initial (pre-expansion) radial diffusion timescale.  Condition~(\ref{eq:L_cj_heirarchy_condition}) is equivalent to a range of break-out times $0.4 \, {\rm d} < t_{\rm bo} < 4.2 \, {\rm d}$ according to equation~(\ref{eq:tbo}), somewhat longer than those expected for SLSNe engines.

In this timescale hierarchy, radiation leaks azimuthally as the cocoon emerges from the ejecta.  The light-curve duration and luminosity are approximately given by
\begin{align}
&t_{\rm pk} \approx t_{\rm exp,0} = \frac{v_{\rm ej} t_{\rm bo}}{\beta_{\rm c,j} c}
\sim 2 \times 10^3 - 2 \times 10^4 \, {\rm s} ,
\\
&L_{\rm pk} \approx \frac{E_{c,j}}{t_{\rm exp,0}} \approx f_{\Gamma\beta,1} \frac{\beta_{\rm c,j} c}{v_{\rm ej}} L_{\rm j}
\sim 10^{44} - 2 \times 10^{45} \, {\rm erg \, s}^{-1} .
\end{align}

In summary, the most promising source of isotropic cocoon emission originates from the part of the shocked-jet material that attains transrelativistic velocities following jet break-out.  Thermal energy released as this matter expands from the jet head may power a UV flare, which lasts a few hours starting after jet break-out (which itself is typically delayed by hours or more after the explosion) and reaches a peak luminosity of $\gtrsim 10^{44}$ erg s$^{-1}$.

\subsection{Thermal Wind from Jet-Ejecta Interface} \label{subsubsec:jet_energized_emission}
\label{sec:jetwind}

While a potentially promising target for future high cadence surveys like the Zwicky Transient Facility (ZTF; \citealt{Bellm14}) and ULTRASAT \citep{Ganot+16}, the cocoon break-out emission described in the previous section cannot explain the early maxima observed in SLSNe light curves (\citealt{Leloudas+12,Nicholl+15b,Nicholl&Smartt16,Smith+16}).  The latter instead occur over a duration of several days, a timescale which is suggestively comparable to the lifetime $t_{\rm e}$ of the engine needed to power the peak of the SLSNe at later times.  

One explanation for this coincidence is that the early emission is powered by a spherical shock break-out through the ejecta, driven by the pressure of the pulsar wind nebula inflated by the magnetar wind \citep{KasenMetzger&Bildsten16,Chen+16,Suzuki&Maeda16}.  However, \citet{KasenMetzger&Bildsten16} found that it was challenging to make this emission component stick out above the rising SN light curve unless the thermalization efficiency of the engine power is suppressed at early times.  Here we outline a new mechanism for producing early-time emission from an off-axis jet, also acting on a timescale comparable to the engine duration $t_{\rm e}$.

Even after the jet has breached the stellar surface, and once a steady outflow of relativistic matter has been established, the jet should continue to transfer a small fraction of its kinetic energy and momentum to the confining ejecta walls.  Common sense dictates that the process by which a jet escapes the star will not be entirely ``clean", a fact which should have observable consequence for an off-axis observer.  The process by which this ``friction'' occurs is left unspecified, but it could plausibly be related to intrinsic variability of the central engine \citep{Morsony+10,Lazzati+11}, conical recollimation shocks between the jet and confining (evacuated) funnel \citep[e.g.][]{BarniolDuran+16}, Kelvin-Helmholtz instabilities at the jet-ejecta interface, or some other unspecified process.  High resolution numerical hydrodynamical simulations, focusing on this shearing interface, are needed to quantify the efficiency of this energy deposition and the mass entrained by this process.  

To make progress, in the following we simply assume the jet deposits heat into the ejecta walls at a rate of $\dot{E}_{\rm w} = 2 \epsilon L_{\rm j}$ (where $\epsilon \ll 1$), driving mass-loss at a rate $\dot{M}_{\rm w}$.  We find it convenient to express the mass and energy loss rates in terms of the terminal wind velocity (achieved after this heat is reconverted into kinetic energy via adiabatic expansion) according to 
$v_{\rm w} = (2 \dot{E}_{\rm w} / \dot{M}_{\rm w})^{1/2} $
because the jet interface-driven wind scenario is only meaningful across a small range of parameters, $v_{\rm ej} \lesssim v_{\rm w} < c$.  

The characteristic rise timescale for the light curve produced by this heated wind is again set by radiative diffusion,
\begin{align} \label{eq:t_pk_wind}
t_{\rm pk} = \frac{3 \kappa \dot{M}_{\rm w}}{4 \pi c v_{\rm w}}
&= \frac{3 \kappa \epsilon L_{\rm j}}{\pi c v_{\rm w}^3} 
\approx 1.4 \, {\rm d} \, \left(\frac{\kappa}{0.2 \, {\rm cm}^2 \, {\rm g}^{-1}}\right)
\\ \nonumber 
&\times \left(\frac{2 \epsilon L_{\rm j}}{10^{45}\,{\rm erg \, s}^{-1}}\right) \left(\frac{v_{\rm w}}{3 \times 10^{9} \, {\rm cm \, s}^{-1}}\right)^{-3}
.
\end{align}
After this time, the {\it bolometric} luminosity will track the engine (jet) power,
\begin{equation} \label{eq:L_wind_pk}
L(t_{\rm bo}+t_{\rm pk} \lesssim t) \approx \dot{E}_{\rm w}(t) 
= 2 \epsilon L_{\rm j}(t),
\end{equation}
consistent with the standard \citet{Arnett79} rule.

Across the range of radii $v_{\rm ej}t \ll r \ll R_{\rm bo}+v_{\rm w} (t-t_{\rm bo})$, 
the wind assumes a secularly-evolving `steady-state' density profile where $\rho_{\rm w}(t) \approx \dot{M}_{\rm w}(t) / 4\pi v_{\rm w} r^2$, and
the radius of the photosphere is approximately given by
\begin{align}
R_{\rm eff} &= \frac{\kappa_{\rm es} \dot{M}_{\rm w}(t) }{ 4 \pi v_{\rm w}} 
= \frac{\kappa_{\rm es} \epsilon L_{\rm j}(t)}{\pi v_{\rm w}^3} 
\nonumber \\
&\simeq 1.2 \times 10^{15} \, {\rm cm} \, 
\left(\frac{2 \epsilon L_{\rm j}(t)}{10^{45}\,{\rm erg \, s}^{-1}}\right) \left(\frac{v_{\rm w}}{3 \times 10^{9} \, {\rm cm \, s}^{-1}}\right)^{-3} ,
\end{align}
where we have adopted $\kappa_{\rm es} = 0.2$ cm$^{2}$ g$^{-1}$ as the electron scattering opacity.
The effective temperature of the emission is given by
\begin{align} \label{eq:Teff_wind}
T_{\rm eff} &\equiv \left(\frac{L}{4\pi \sigma R_{\rm eff}^2}\right)^{1/4} 
= \left(\frac{\pi v_{\rm w}^6}{2 \sigma \kappa^2 \epsilon L_{\rm j}}\right)^{1/4}
\\ \nonumber
&\simeq 3.2\times 10^{5} \,{\rm K}\,\left(\frac{2 \epsilon L_{\rm j}(t)}{10^{45}\,{\rm erg \, s}^{-1}}\right)^{-1/4} \left(\frac{v_{\rm w}}{3 \times 10^{9} \, {\rm cm \, s}^{-1}}\right)^{3/2} ,
\end{align}
where we have expressed our results in terms of $2 \epsilon L_{\rm j}(t)$, the peak bolometric wind luminosity (equation~\ref{eq:L_wind_pk}).

The high temperatures implied by equation~(\ref{eq:Teff_wind}) show that the interface wind emission is typically observed on the Rayleigh-Jeans tail. The optical-band luminosity in this regime scales as $\propto L_{\rm bol} T_{\rm eff}^{-3} \propto L_{\rm bol}^{1/4} R_{\rm eff}^{3/2}$, and therefore depends sensitively on emission temperature.  As the engine luminosity decays, the radius of the expanding ejecta will overtake the photosphere of the interface wind.  Once $v_{\rm ej} t \gtrsim R_{\rm eff}(t)/2$, the optically-thick spherical wind model described above breaks down, and the wind radiates immediately after emerging from the ejecta.  The effective emitting region shrinks to the (one-sided) `polar cap', $R_{\rm eff} \sim 0.5 \theta_{\rm j} v_{\rm ej} t$, causing an abrupt drop in the optical-band luminosity, primarily due to the rapidly rising temperature of the wind emission.\footnote{Note that once the wind becomes optically thin, the emission reaching a typical off-axis observer will originate from only one side of the bi-polar jets; therefore the bolometric luminosity in this regime is a factor of two smaller than implied by equation~(\ref{eq:L_wind_pk}).}

\begin{figure}
\includegraphics[width=0.5\textwidth]{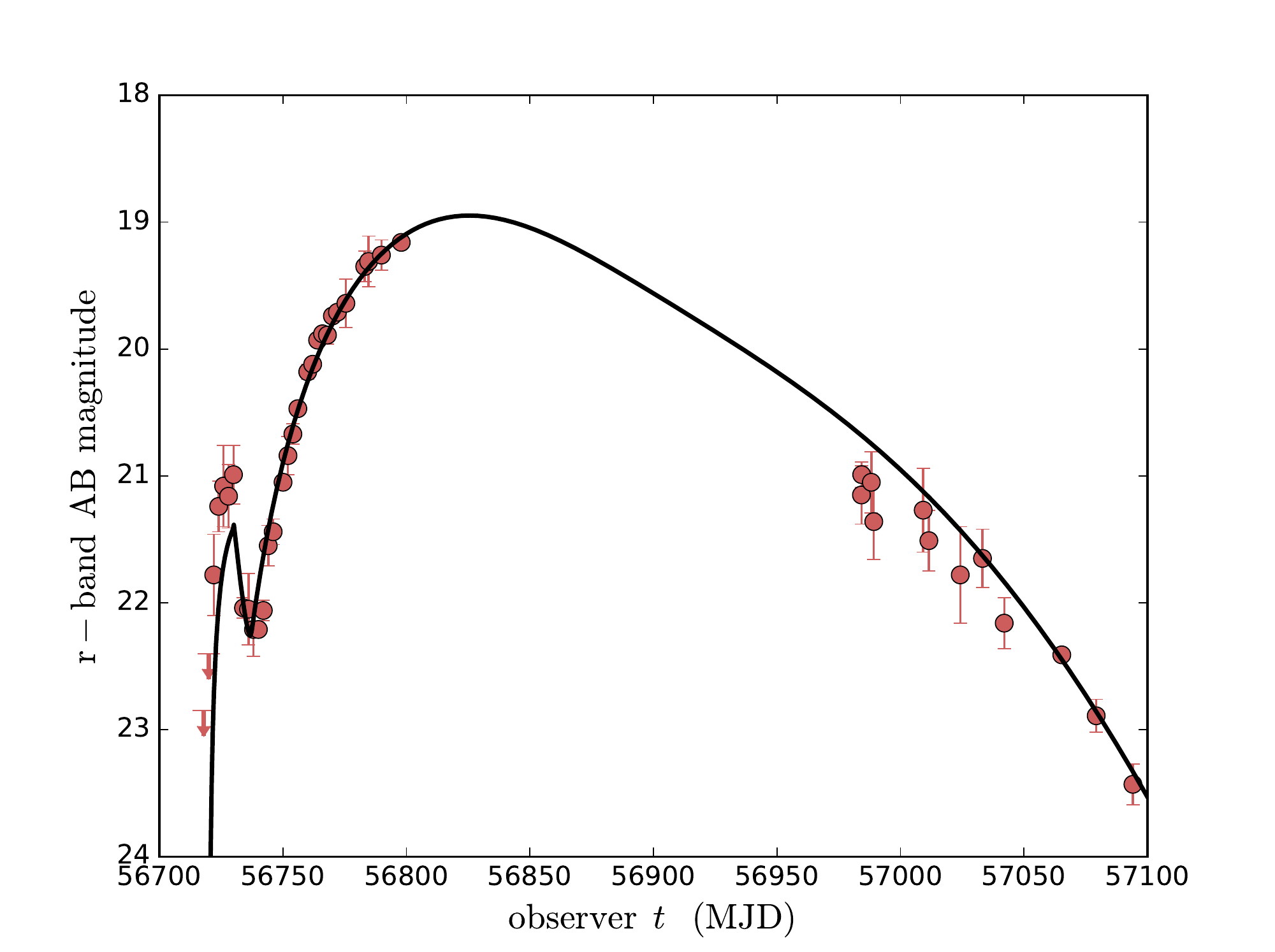}
\caption{Model fit to the r-band light curve (solid black line) of the double peaked SLSN LSQ14bdq (red points).  The main SN peak is powered by radiative diffusion through the ejecta, as in the standard engine-powered supernova model.  The early maximum is instead thermal emission powered by a hypothesized wind driven from the interface between the off-axis relativistic jet (which has successfully escaped from the star) and the supernova ejecta ($\S\ref{sec:jetwind}$).  The best-fit model parameters are: $E_{\rm e}=7.4 \times 10^{52} \, {\rm erg}$, $t_{\rm e}=37.2 \, {\rm day}$, $M_{\rm ej}=5.3 M_\odot$, $E_{\rm sn}=3.3 \times 10^{50} \, {\rm erg}$, $f_{\rm j}=0.55$, $\epsilon=0.14$, and $v_{\rm w}=3.2 \times 10^9 \, {\rm cm \, s}^{-1}$.
}  \label{fig:lightcurve}
\end{figure}

Figure~\ref{fig:lightcurve} shows a fit of our engine-powered SN + interface wind model to the r-band light-curve of the double-peaked SLSN LSQ14bdq \citep{Nicholl+15}. The model can reproduce both the early bump and main SN peak within the same framework, for a reasonable set of parameters as marked in the caption.  The numerical model from which the fit is derived is calculated by integrating the two-zone SN+wind equations while self-consistently evolving $R(t)$, $E(t)$ for each component, as described in Appendix~\ref{sec:Appendix_B}. The effective photospheric radius of each component is tracked at each timestep, from which the observed optical-band luminosities are then calculated assuming thermal emission.
In particular, the SN photosphere is calculated by integrating the density profile equation~(\ref{eq:rho_ejecta}) and assuming a constant opacity, resulting in
\begin{equation}
R_{\rm eff, sn}(t) =
v_{\rm ej} t
\begin{cases}
\exp \left[ - \frac{v_{\rm ej}}{c} \left(\frac{t}{t_{\rm pk, sn}}\right)^2 \right] &; \, \delta=1 \\
\left[ 1 + \frac{(\delta-1) v_{\rm ej}}{c} \left(\frac{t}{t_{\rm pk, sn}}\right)^2 \right]^{-1/(\delta-1)} &; \, \delta \neq 1
\end{cases} ,
\end{equation}
where $t_{\rm pk, sn} = \sqrt{\zeta_\rho \kappa M / v_{\rm ej} c}$.
We note that the simple estimate above neglects temperature and ionization effects on the ejecta opacity, which may become increasingly important at late times $t \gg t_{\rm pk, sn}$.

\subsection{Orphan Radio Afterglow --- the Case of SN2015bn} \label{sec:RadioAfterglow}
Our finding that energetic off-axis jets may commonly accompany SLSNe also predicts radio afterglows from these events produced by the interaction between the relativistic jet and the surrounding external medium.  Though not as bright at early times as the radio emission for an on-axis observer, off-axis viewers should nevertheless observe radio emission once the jet decelerates to mildly relativistic speeds, allowing the observer to enter the causal emission region (a so-called ``orphan afterglow", because the prompt jet emission was relativistically beamed away from the observer line of sight).

Relatively few constraining radio follow-up observations have been published of SLSNe \citep{Chomiuk+11}.  \cite{Nicholl+16} recently obtained deep radio limits on the nearby SN2015bn, which were claimed to rule out a `healthy' jet coincident with this event.  Here we re-examine these radio non-detection constraints
in greater detail
by conducting a parameter survey over the two primary afterglow parameters: the injected jet energy $f_{\rm j} E_{\rm e}$ and the ambient density. The latter is parameterized by $n_0$ for the case of a constant density interstellar-medium (ISM) surrounding, and the wind mass-loss parameter $A_\star$ for the case of a wind environment,
\begin{equation}
\rho_{\rm CSM}(r) = 
\begin{cases}
n_0 m_{\rm p} &; \, {\rm ISM}
\\
5\times10^{11} \, {\rm g \, cm}^{-1} \, A_\star r^{-2} &; \, {\rm Wind}
\end{cases}
\end{equation}
where $A_\star = 1$ corresponds to a progenitor stellar wind of velocity $10^{3}$ cm s$^{-1}$ and mass loss rate $10^{-5} \, M_\odot \, {\rm yr}^{-1}$.

We adopt a semi-analytic afterglow model. The blast-wave dynamics are calculated following \cite{Oren+04} and assuming azimuthal expansion of the jet at the local sound speed \citep[see also][]{Huang+00}, while the emitted synchrotron radiation is calculated following \cite{Sari+98}. We neglect self-absorption, which is irrelevant at these late epochs, and adopt conservative fiducial values of 
$\theta_{\rm j} = \epsilon_e = \epsilon_B = 0.1$, $p=2.5$
for the jet half-opening angle and microphysical parameters. Smaller values of $\theta_{\rm j}$, $\epsilon_e$ or $\epsilon_B$ would only reduce the model-predicted flux and thus weaken  the non-detection constraints.
Following a procedure similar to \cite{Soderberg+06},
we survey the jet-energy -- ambient-density parameter space and locate `allowed' regions, where the model-predicted $22 \, {\rm GHz}$ ($7.4 \, {\rm GHz}$) radio flux at $t_{\rm obs} = 330 \, {\rm day}$ falls below the $40 \, \mu{\rm Jy}$ ($75 \, \mu{\rm Jy}$) upper limit constraints on SN2015bn. In practice, we find the $7.4 \, {\rm GHz}$ upper limit unconstraining in light of the deeper $22 \, {\rm GHz}$ limits (see also \citealt{Nicholl+16}).

\begin{figure}
\centering
\includegraphics[width=0.5\textwidth]{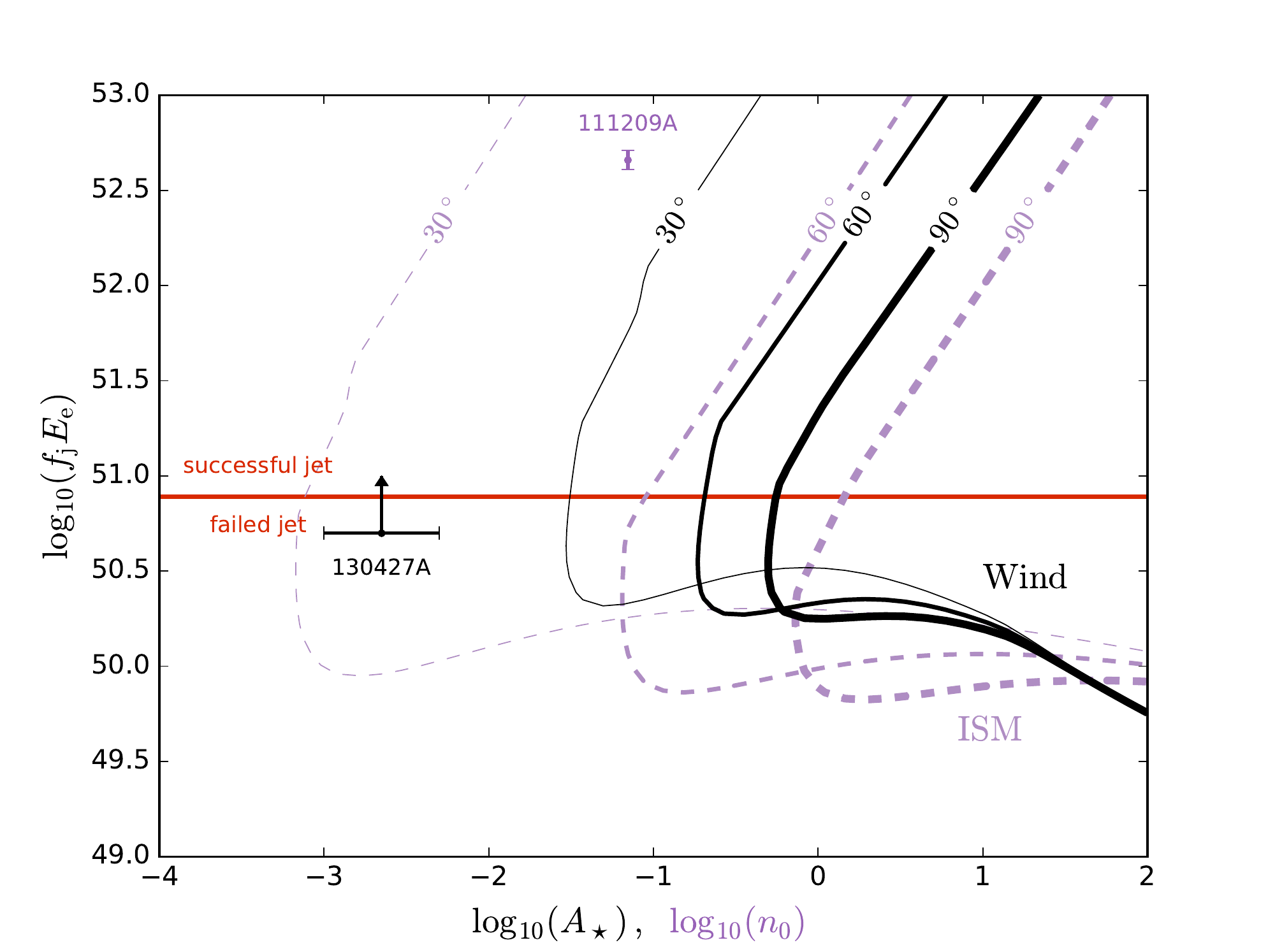}
\caption{Constraints on jet energy and ambient density for SN2015bn, based on late time radio non-detections from \citet{Nicholl+16}. 
Solid black (dashed purple) curves separate allowed and forbidden regions of parameter-space  for different angle off-axis observers, assuming a wind (ISM) ambient-density profile.
Regions to the left (lower-density) of any given curve are permitted.
The horizontal red curve shows the approximate condition on late-time jet breakout (equation~\ref{eq:Ee_min}). 
Also shown are best-fit parameters from detailed afterglow modelling of the ULGRB 111209A and GRB 130427A \citep{Stratta+13,Perley+14}.
A reasonably powerful jet accompanying SN2015bn cannot be ruled out for most off-axis observers if this event occurred in an environment similar to GRB 130427A and GRB 111209A.
} \label{fig:2015bn_radio}
\end{figure}

Fig.~\ref{fig:2015bn_radio} shows our results for both constant density (dashed purple curves) and wind (solid black curves) environments. 
Heavy, medium, and lightly weighted curves show constraints for off-axis observers at viewing angles of $90^\circ$, $60^\circ$, and $30^\circ$ respectively.
The regions to the right of each curve are ruled-out for a given observer.
The horizontal red curve shows the jetted outflow's required energy for successful breakout  (equation~\ref{eq:Ee_min}). We therefore do not expect successful jets below this limit.

Even with this additional constraint, we find that there is a significant parameter-space where a `healthy' jetted outflow coincident with SN2015bn could have gone undetected.
While this would require a low ambient density, such densities are in fact inferred from observations of many LGRB afterglows, as indicated by the points in the plot.
For example, best-fit models of the extraordinarily well-observed afterglow of GRB 130427A \citep{Perley+14} yield $A_\star \sim 10^{-3}$. Similarly, afterglow modeling of the ULGRB 11209A associated with SLSN 2011kl implies an ISM density of $n_0 = 0.07 \, {\rm cm}^{-3}$ \citep{Stratta+13}.
We conclude that a relativistic jet accompanying SN2015bn would go undetected by most off-axis observers if it had occurred in a similar environment to these well-studied events.

A low-density external environment for SN2015bn is also consistent with X-ray upper limits recently obtained by \cite{Margutti+17}, which imply $A_\star < 10^3$. 
Tighter constraints of $A_\star < 2$ are obtained by these authors for the SLSN PTF12dam.
More broadly, for the SLSN-I population as a whole, \cite{Margutti+17} find that energetic jets are not constrained by current X-ray data for off-axis observers at significant viewing angles $\gtrsim 30^\circ$.
We therefore conclude that current upper-limits cannot rule-out a reasonably powerful jet coincident with the SLSN 2015bn, as we might expect would have accompanied it.

\section{Discussion and Implications} \label{sec:Discussion}

\begin{figure*}
\centering
\includegraphics[width=0.95\textwidth]{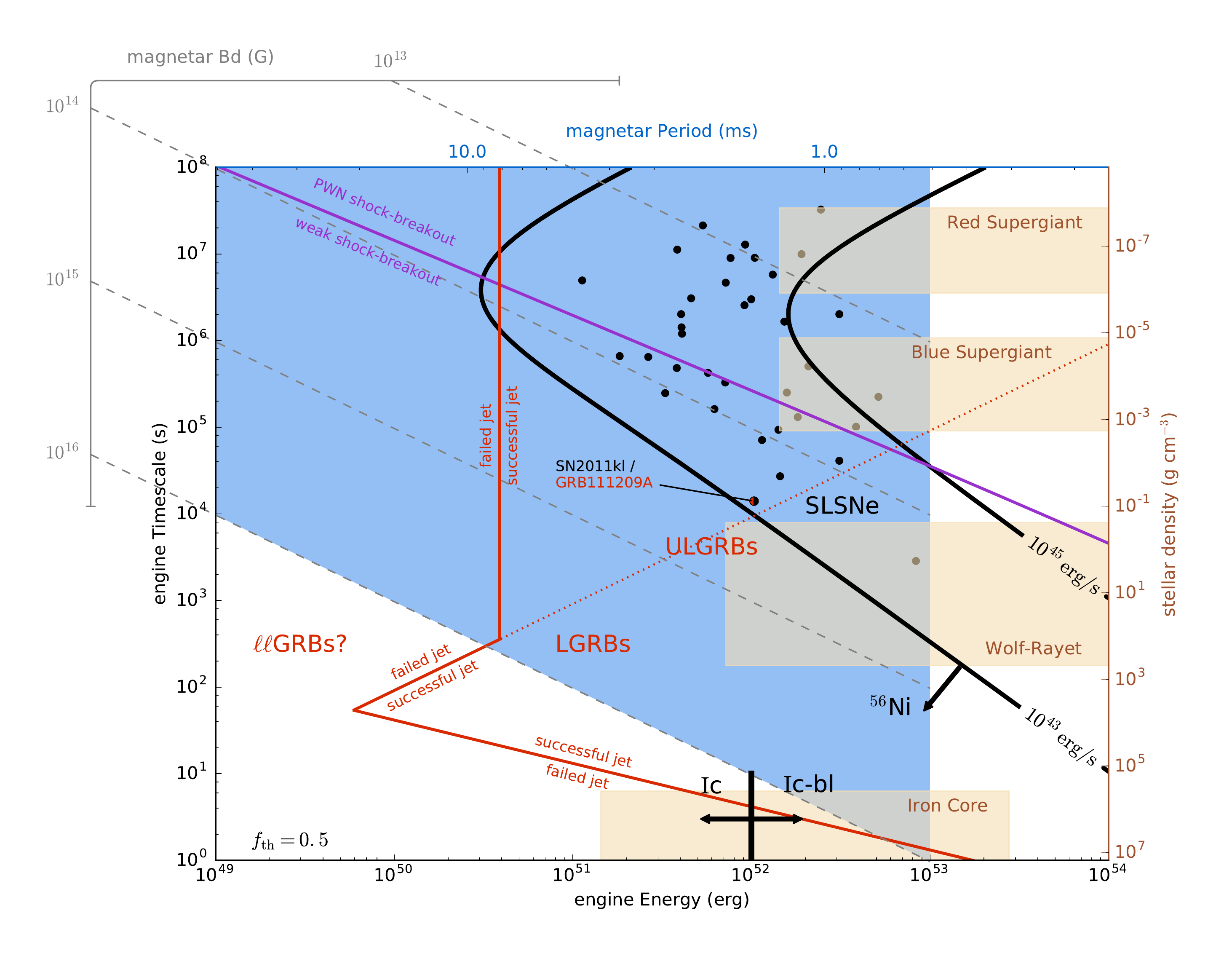}
\caption{
Central-engine phase space, plotted for equal thermal and jetted energy fractions ($f_{\rm th}=f_{\rm j}=0.5$, equivalent to $\alpha \simeq 0.4$ in the magnetar model; Fig.~\ref{fig:fmfth}): 
the generic axes of engine timescale $t_{\rm e}$ and energy $E_{\rm e}$ are related to magnetar spin-period $P_0$ and dipole magnetic field $B_{\rm d}$ in the magnetar scenario (equations~\ref{eq:Ee_magnetar},\ref{eq:te_magnetar}), or average density of the stellar progenitor and fall-back mass $E_{\rm e}/\epsilon_{\rm fb} c^2$ in accretion-powered model. The blue-shaded region bounds the range of plausible magnetar parameters ($E_{\rm e} \lesssim 10^{53} \, {\rm erg}$, $B_{\rm d} \lesssim 10^{16} \, {\rm G}$) and light brown shaded regions depict several fall-back progenitor model parameters.  Black contours show the peak SN luminosity powered by the central engine, while black points show the best-fit parameters of a population of observed
Type I SLSNe. 
The solid red curve separates between parameter-space regions where a self-collimated jet manages or fails to break-out of the entraining stellar matter (equations~\ref{eq:te_min},\ref{eq:Ee_min}). 
Above the dotted red curve (equation~\ref{eq:Lstar_bo_tstar}), such break-out occurs within the expanding SN ejecta.
The figure illustrates the diversity of transients which can arise from the collapse of rapidly rotating stellar cores, and shows that successful jets may commonly accompany SLSNe.
Adopted model parameters are $\gamma_{\rm j}=2$, $R_{\star} = 10^{11}$ cm, $E_{\rm sn} = 10^{51} \, {\rm erg}$, $M_{\rm ej} = M_\star = 5 M_\odot$, and for the SN luminosity contours we adopt ejecta opacity $\kappa = 0.2 \, {\rm cm}^2 \, {\rm g}^{-1}$ and central engine power-law decay rate $\ell = 2$ (equation~\ref{eq:Le}).
See text for further details.
} \label{fig:ParameterSpace}
\end{figure*}

We have explored implications of the growing observational connection between long GRB and SLSNe, and their likely common association with the birth of energetic compact objects. 
Both phenomena can conveniently be interpreted within a single theoretical framework of engine-powered transients (equation~\ref{eq:Le}; \citealt{Metzger+15}).
We have focused explicitly on magnetar engines, for which we have proposed a novel mechanism of driving both jetted and thermal outflows (\S\ref{sec:MagnetarModel}), but our subsequent results for weak-jet break-out (\S\ref{sec:Jet}) and associated observational signatures (\S\ref{subsec:JetSignature}) can equally be applied to alternative engine models within this framework (e.g. the black-hole accretion, i.e. `fall-back', scenario).

\subsection{Landscape of Engine-Powered Transients}

Figure~\ref{fig:ParameterSpace} summarizes the landscape of such engine-powered transients, and illustrates the diversity of potential observational signatures which may be produced by the collapse of rapidly rotating stellar cores, including long GRB, ULGRB, broad-lined SNe-Ic, and SLSNe-I.
Furthermore, it demonstrates the result of equation~(\ref{eq:Ee_min}) --- that SLSNe may generically be accompanied by successful jets.

The primary axes of Fig.~\ref{fig:ParameterSpace} are the total energy $E_{\rm e}$ and characteristic lifetime $t_{\rm e}$ of the engine.  In this example the energy is partitioned equally between the jetted and thermal components, i.e. $f_{\rm th} = 0.5$, corresponding to a dipole inclination angle $\alpha \approx 0.4$ in the magnetar model (Fig.~\ref{fig:fmfth}).  In the magnetar model, the values of $E_{\rm e}$ and $t_{\rm e}$ are related to the dipole field $B_{\rm d}$ and initial spin period $P_0$ through equations~(\ref{eq:Ee_magnetar},\ref{eq:te_magnetar}), which we show as additional axes floating above the top of the figure and which cross the diagram as inclined lines.  The blue shaded region highlights the rather generous parameter space spanned by realistic constraints on the magnetar engine based on the maximum rotational energy $E_{\rm e} \lesssim 10^{53} \, {\rm erg}$ (corresponding to centrifugal break-up for a massive neutron-star; \citealt{Metzger+15}) and maximum realistic magnetic field strength $B_{\rm d} \lesssim 10^{16} \, {\rm G}$.  In the black hole accretion model, the engine energy and lifetime are instead related, respectively, to the total accreted mass ($\sim E_{\rm e} / \epsilon_{\rm fb} c^2$) and mean density of the stellar core, the latter being shown in a separate axis to the right of the figure.  Light brown regions denote the space of fall-back accretion models associated with different stellar progenitor types 
(iron core, Wolf-Rayet outer layers, and Red/Blue Supergiant envelopes, as labeled; \citealt{Sukhbold&Woosley14,Sukhbold+16}), where a span in $\epsilon_{\rm fb} = 10^{-3}-10^{-1}$ is taken.

Black annotations in Fig.~\ref{fig:ParameterSpace} relate to the SN emission.  Black curves depict contours of constant peak SN luminosity $L_{\rm e,pk}$,\footnote{We consider only engine-powered emission, neglecting any contribution from $^{56}$Ni.} with black points showing the engine properties required to explain individual observed SLSNe within the magnetar model (Nicholl et al.~in preparation).   The SLSNe populating this region require a central engine which deposits $\sim 10^{52}\,{\rm erg}$ of {\it thermal} energy over an engine lifetime of typically several days, corresponding to $P_0 \gtrsim 1\,{\rm ms}$ and $B_{\rm d} \sim 10^{14}\,{\rm G}$ in the magnetar scenario.  A black-and-red point shows the best fit central-engine model of \cite{Metzger+15} to SN 2011kl and the associated ULGRB 111209A \citep{Greiner+15}, which straddles the region between typical SLSNe and observed ULGRB and thus represents a potential hybrid or transitional event (see also \citealt{Ioka+16,Gompertz&Fruchter17}).

SNe which are instead powered predominantly by $^{56}$Ni occupy the region to the bottom left of the $L_{\rm e,pk}=10^{43}\,{\rm erg \, s}^{-1}$ contour.  In this region, the engine duration is too short to appreciably contribute to the SN luminosity: the majority of the energy deposited by the engine suffers adiabatic degradation by the time the ejecta becomes transparent around the time of the SN peak. The engine can still enhance the kinetic energy of the ejecta in this case, as long as $f_{\rm th} E_{\rm e} \gtrsim E_{\rm sn}$, where $E_{\rm sn}$ is the initial kinetic energy of the explosion.  Standard Ni-powered SNe Ic are therefore divided from their energetic broad-lined counterparts (`hypernova') at approximately $E_{\rm e} \sim 10^{52} \, {\rm erg}$.\footnote{However, note that kinetic energy is a challenging quantity to measure accurately from the supernova spectra if the ejecta are highly asymmetric (e.g., \citealt{Dessart+17}).}

Thermal energy deposited by the magnetar inflates a hot bubble behind the SN ejecta, analogous to a pulsar wind nebula.  \citet{KasenMetzger&Bildsten16} show that, for a sufficiently energetic engine, this hot bubble drives a shock wave through the outer layers of the SN ejecta, producing an early-time shock-break out signature (see \citealt{Chen+16,Suzuki&Maeda16} for two-dimensional simulations of this process).  The purple curve shows the combination of energy and lifetime above which this shock break-out emission is particularly pronounced, because it occurs while the engine is still near its peak activity (corresponding to quadrants 1 and 2 of Fig.~2 of \citealt{KasenMetzger&Bildsten16}).  This break-out was proposed by \citet{KasenMetzger&Bildsten16} as a explanation for the early peak observed in the light curves of SLSNe \citep{Leloudas+12,Nicholl+15,Nicholl&Smartt16,Smith+16}, to which we have offered an alternative explanation (\S\ref{subsubsec:jet_energized_emission}).

Red curves and annotations within Fig.~\ref{fig:ParameterSpace} relate to the jetted component of the engine, and show the approximate regions of parameter space giving rise to classical GRB, ULGRB, and low-luminosity GRB ($\ell \ell$GRB).  A red curve separates the engine energy above which the relativistic jet successfully escapes the SN ejecta instead of being choked (equations~\ref{eq:te_min},\ref{eq:Ee_min}). 
This criterion is a non-trivial function of the engine properties because --- in the case of a long engine duration --- the ejecta has time to expand appreciably following the explosion (\S\ref{subsec:BreakoutConditions}). Even relatively weak jets 
below a critical luminosity $\simeq 3 \times 10^{47} \, {\rm erg \, s}^{-1}$ (dotted red curve; equation~\ref{eq:Lstar_bo_tstar}),
which could not escape the (effectively stationary) stellar progenitor in the case of a short-lived engine, are capable of escaping at late times as the ejecta dilutes following its expansion in the SN explosion. 

Figure~\ref{fig:ParameterSpace} therefore summarizes the break-out criterion derived in \S\ref{subsec:BreakoutConditions}. Importantly, it illustrates that observed SLSNe inhabit
a parameter-space region where they are expected to be accompanied by successful, albeit low-luminosity, jets. 
Although we have assumed $f_{\rm j}=0.5$ in plotting Fig.~\ref{fig:ParameterSpace}, this result is robust and holds for even modest jetted energy fractions of only several percent.
We thus conclude that off-axis jets may be a common feature of SLSNe.

\subsection{Jetted High-Energy Transients}

The early break-out times of jets from SLSNe engines (within days of the explosion) implies that their signatures may be missed in most optically-selected events, which are generally discovered at later times and are viewed off the jet axis.  

\begin{figure}
\centering
\includegraphics[width=0.5\textwidth]{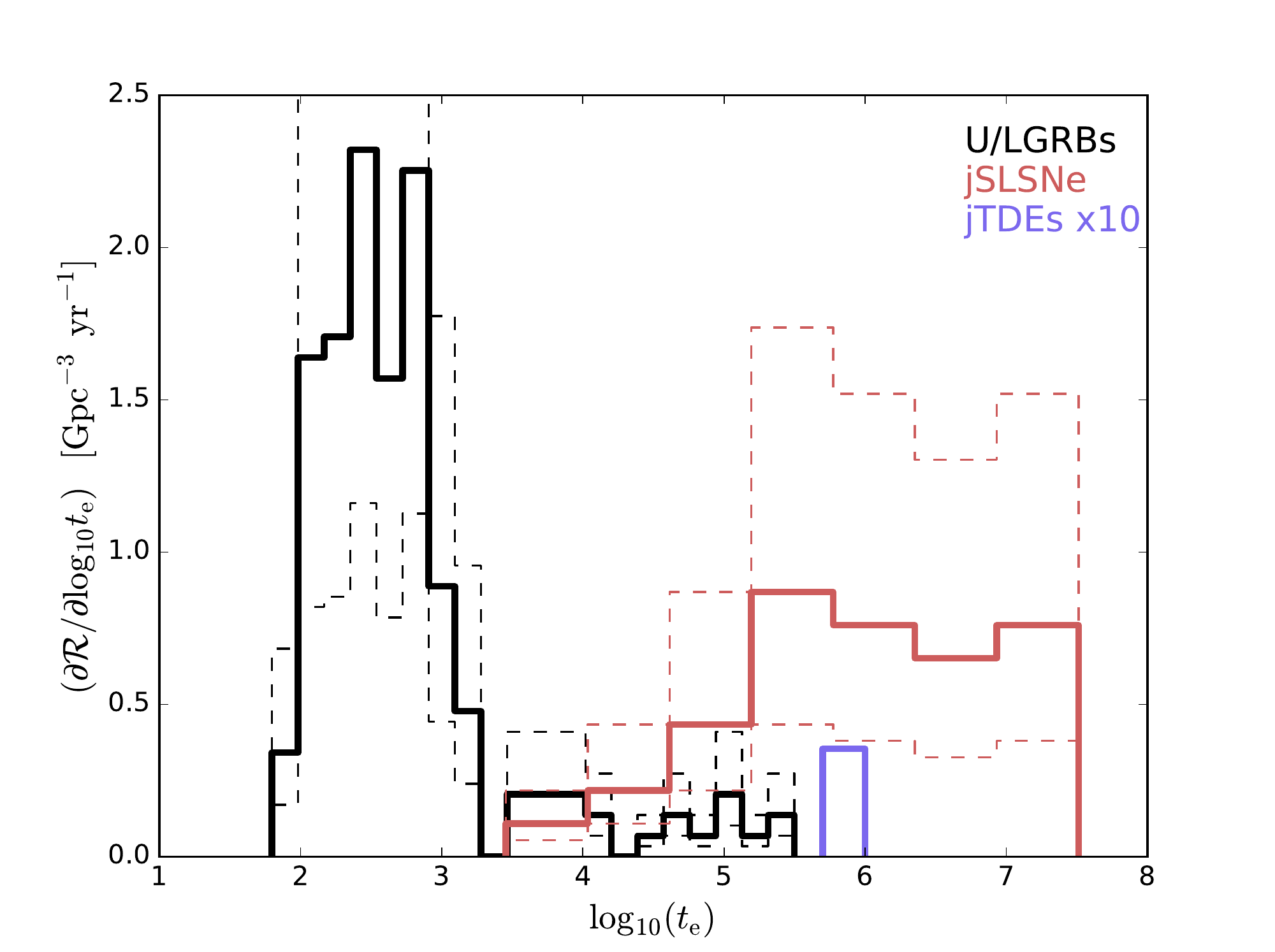}
\includegraphics[width=0.5\textwidth]{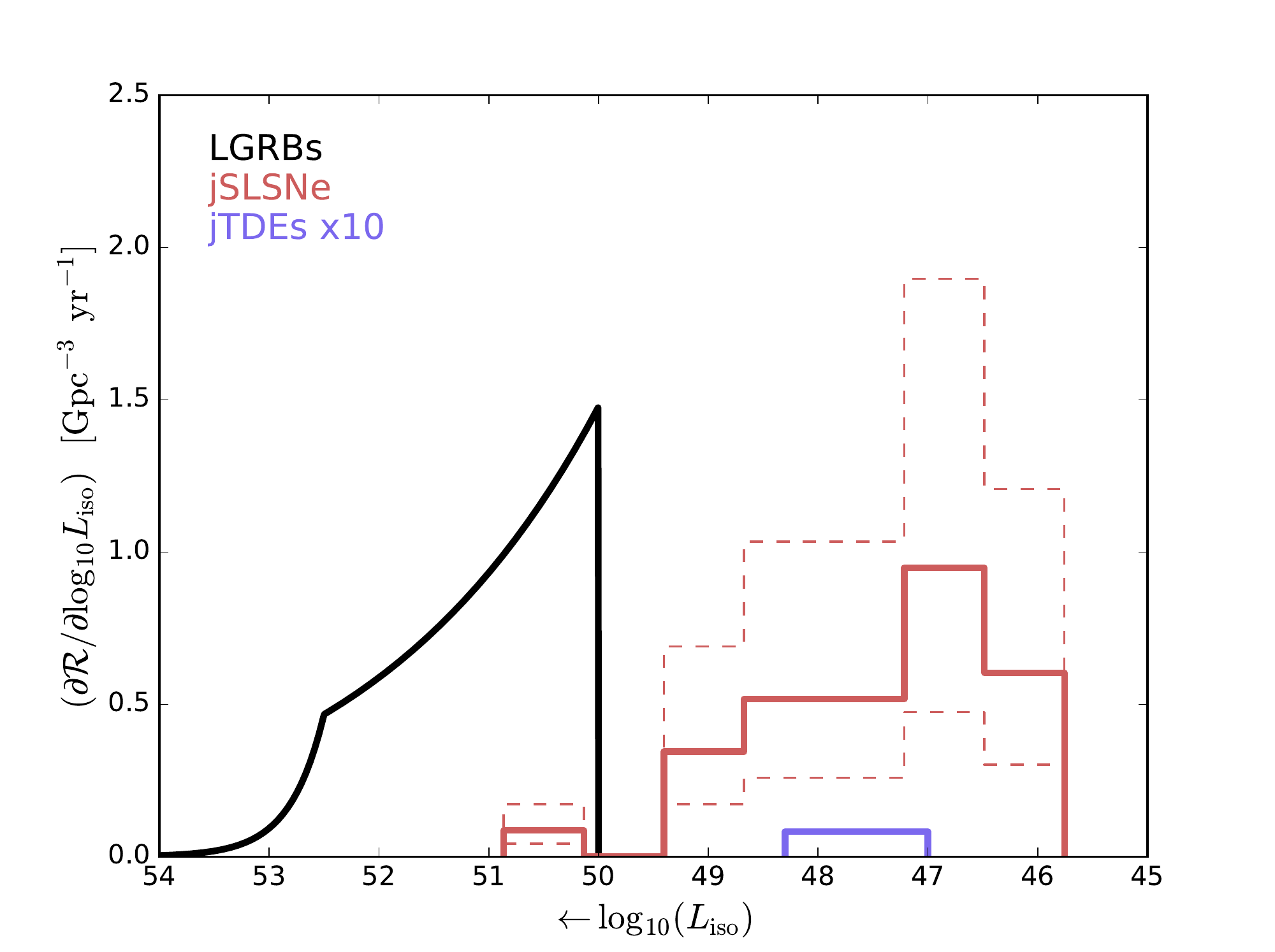}
\caption{Distribution of duration $t_{\rm e}$ (in seconds) and peak isotropic luminosity $L_{\rm iso}$ (in erg s$^{-1}$) of jetted transients, normalized to their volumetric rates at redshift $z \lesssim 1$.
{\it Top Panel}: Measured distribution of U/LGRB engine durations (\citealt{Zhang+14}; solid-black) and jetted-tidal disruption events (blue; scaled up by a factor of $\times 10$) compared to the predicted durations distribution of jets accompanying SLSNe (solid-red; based on the engine duration obtained by fitting the magnetar model to the SLSNe optical light curves).  Dashed curves account for an assumed factor of two uncertainty in the event rates.
{\it Bottom Pannel}: Isotropic-equivalent luminosity function of long GRB (as derived by \citealt{Wanderman&Piran10}, accounting for detection bias; solid-black) and jetted tidal disruption events (blue; scaled up by a factor of $\times 10$) compared to the predicted distribution for jets accompanying SLSNe  (solid-red).  Note that luminosity decreases to the right, facilitating an easier comparison with the duration distribution.
} \label{fig:jets}
\end{figure}

However, if successful jets are commonly associated with SLSNe, one implication is the existance of a class of high energy transients with extremely long durations $T_{\rm 90} \sim 10^{5}-10^{6}$ s (e.g.~\citealt{Quataert&Kasen12,Woosley&Heger12}) and associated bright optical counterparts (\citealt{Metzger+15}).  Figure \ref{fig:jets} shows our predictions for the duration and peak luminosity distribution of these SLSNe-associated events, which we have calculated using the engine properties (energy $E_{\rm e}$, lifetime $t_{\rm e}$, assuming a jet fraction $f_{\rm j} = 0.5$) and volumetric rates of the observed sample of SLSNe (\citealt{Quimby+13}), and assuming a beaming fraction of $f_{\rm b} = 1/50$, similar to those typical of long GRBs. 
The density is normalized with respect to the average volumetric event rate of such jetted events within $z \leq 1$ (following a calculation similar to equation~\ref{eq:Ndot_ULTRASAT}), so that the integral over the distribution equals $\left\langle \mathcal{R}_{\rm jSLSN} \right\rangle \simeq 2.2 \, {\rm Gpc}^{-3} \, {\rm yr}^{-1} \, \left( f_{\rm b}^{-1} / 50 \right)^{-1}$. 

Shown for comparison is the engine duration distribution 
of the observed long and ultra-long GRBs, adapted from \citet{Zhang+14},
and the inferred LGRB luminosity function from \cite{Wanderman&Piran10} (similarly normalized with respect to the $z \leq 1$ GRB rate, $\left\langle \mathcal{R_{\rm LGRB}} \right\rangle \simeq 2.3 \, {\rm Gpc}^{-3} \, {\rm yr}^{-1}$).
Our estimates show that the predicted rate of jetted transients from SLSNe is comparable to the 
long GRB rate, though the distribution is bimodal.  This deficit of intermediate luminosity and duration jets may be physical and related to a bimodal population of engine properties, or it may be the result of selection effects against detecting long-lived low-luminosity gamma-ray transients (\citealt{Gendre+13,Levan+14}) or of identifying engine-powered supernovae when their engines have shorter durations and they may not be classified as ``superluminous''.  

The properties of the SLSNe-associated jets we predict are similar to those of the ``jetted tidal disruption events'' (\citealt{Bloom+11,Levan+11,Burrows+11,Zauderer+11}), of which there are currently only three examples \citep{Cenko+12,Brown+15}.  
A lower limit on their rates can be crudely estimated by considering that {\it Swift} has observed $3$ jetted-TDEs, which would have been detected up to redshift $\sim 1$, over a $\sim 12 \,{\rm yr}$ baseline. Therefore $\left\langle \mathcal{R}_{\rm jTDE} \right\rangle \sim f_\Omega (3 \, {\rm events} / 12 \, {\rm yr}) / V(z \leq 1) \simeq 0.01 \, {\rm Gpc}^{-3} \, {\rm yr}^{-1} $, where $f_\Omega \simeq 1/7$ is {\it Swift}'s fractional all-sky field of view, and $V$ the co-moving volume within redshift $z$.
These events are shown for comparison in Fig.~\ref{fig:jets} with blue colored lines.

Although a tidal disruption origin is favored for these high energy transients, in part due to the coincidence of the prototype Swift J1644+57 with the nucleus of its galaxy, this association is not air tight (the possibility of a chance coincidence with the nucleus is high, and the angular size of the host galaxies of the other events are too small to tell), especially considering the lack of evidence for powerful jets in other tidal disruption flares (e.g.~\citealt{Bower+13,vanVelzen+13,Generozov+17}).  This raises the possibility that some or all of these events may in fact be core collapse events (\citealt{Quataert&Kasen12,Levan+14}), possibly of the type described here.  Although Swift J1644+57 was highly dust extincted, it did show evidence for a separate component of (possibly thermal) optical/IR emission \citep{Levan+16}, while Swift J2058+05 showed clear evidence for a very luminous $\sim 10^{45}$ erg s$^{-1}$ thermal component \citep{Cenko+12,Pasham+15}, broadly consistent with the properties of SLSNe.  The fact that the estimated rates of these transients is lower than
those we predict in association with SLSNe could suggest the engine luminosities of the observed ``jetted tidal disruption events" are on the high end of the distribution; that only a small fraction of SLSNe produce successful jets; or that selection effects against detecting long-lived low-luminosity transients (\citealt{Gendre+13,Levan+14}) again result in larger underestimates of the volumetric rates of these transients.

\subsection{UV Flash from Jet Break-Out}

Beyond potential on-axis jet signatures, we have considered other sources of jet-powered emission, which may be visible also for off-axis viewers, i.e.~in coincidence with most/all SLSNe.  One source is UV emission due to the cocoon break-out ($\S\ref{sec:cocoon}$), which is expected to last for a few hours and will reach peak luminosities of $\sim 10^{44}-10^{45} \, {\rm erg \, s}^{-1}$, corresponding to an absolute magnitude $M \sim -21.3$ to $-23.8$.
Cocoon break-out emission from SLSNe could be a promising source for future wide-field UV survey missions, such as ZTF \citep{Bellm14} the proposed ULTRASAT satellite (e.g., \citealt{Ganot+16}).  

ULTRASAT\footnote{\url{http://www.weizmann.ac.il/astrophysics/ultrasat}}  will achieve a sensitivity of 21.5 AB magnitude in the 220-280 nm wavelength range for an integration time of 900 seconds.  Its 210 deg$^{2}$ instantaneous field of view covers a fraction $f_{\Omega} \simeq 5\times 10^{-3}$ of the sky.  
Assuming a detection threshold at twice the sensitivity, 
a source of magnitude $M \sim -21.3$ to $-23.8$ would be visible to $z_{\rm max} \sim 0.4-1.2$, out to which the co-moving volume of the Universe is $V_{\rm max} \sim 23-224$ Gpc$^{3}$.  The rate of SLSNe at $z \approx 0.16$ is estimated to be $\mathcal{R}_{\rm SLSN} \approx 32^{+77}_{-26}$ Gpc$^{-3}$ yr$^{-1}$ (\citealt{Quimby+13}), which if increasing as the star formation rate $\propto (1+z)^{3.28}$ (\citealt{Hopkins&Beacom06}) would increase to $\mathcal{R}_{\rm SLSN} \approx 75-338$ Gpc$^{-3}$ yr$^{-1}$ by $z_{\rm max} \approx 0.4-1.2$.  The number of SLSN jet break-outs detectable by ULTRASAT is therefore very roughly 
\begin{eqnarray} \label{eq:Ndot_ULTRASAT}
\dot{N} &=&  f_{\Omega}\int_{0}^{z_{\rm max}}\frac{\mathcal{R}_{\rm SLSN}(z)}{1+z}\frac{dV}{dz}dz \nonumber \\ &\sim& f_{\Omega}\frac{\mathcal{R}_{\rm SLSN}(z_{\rm max})V_{\rm max}}{1+z_{\rm max}} \sim 5-114\,{\rm yr^{-1}}\ .
\end{eqnarray}

\section{Summary} \label{sec:Conclusions}
This paper investigates the SLSN-GRB connection, exploring the powering-mechanism, break-out conditions, and observational signatures of engine-powered jets.
Our main findings are the following.
\begin{enumerate}

\item Mis-alignment between rotation and magnetic axes of a millisecond magnetar provides a natural mechanism for dissipating a fraction $f_{\rm th}$ of the magnetar's spin-down luminosity and powering an energetic SN (Fig.~\ref{fig:Cartoon}).
The remaining spin-down luminosity, $f_{\rm j} = 1 - f_{\rm th}$, can power an ordered, magnetically dominated jet (Fig.~\ref{fig:fmfth}).
Within this model, $f_{\rm th}$ and $f_{\rm j}$ are solely functions of the mis-alignment angle $\alpha$ (equation~\ref{eq:fth_fit}).
Thus, observational measurements of both jet and SN energies for a common event can be used to infer $\alpha$ (equation~\ref{eq:alpha_fit}).

\item Break-out of weak jets is studied in \S\ref{sec:Jet}, regardless of the underlying mechanism by which they are powered (irrespective of the magnetar model described in \S\ref{sec:MagnetarModel}).
Jets below a threshold luminosity $\simeq 3 \times 10^{47} \, {\rm erg \, s}^{-1}$ (equation~\ref{eq:Lstar_bo_tstar}) cannot break-out of the stellar progenitor before the SN shock-wave reaches the stellar surface (equation~\ref{eq:tstar}).
This regime, which is of interest in the case of SLSNe engines (Fig.~\ref{fig:ParameterSpace}), requires separate treatment from canonical GRB jet models (\S\ref{subsubsec:breakout_from_ejecta}).

\item We find that jet break-out in this regime is set by the condition $v_{\rm h} \gtrsim v_{\rm ej}$ (equation~\ref{eq:vh_bo_criterion}), for which the jet is also kink-stable (equation~\ref{eq:kinkinstability_criterion}).
This yields a simple condition on jet energy for successful break-out, $f_{\rm j} E_{\rm e} \gtrsim 0.19 E_{\rm sn}$ (equation~\ref{eq:Ee_min}). In the magnetar scenario this is commensurate with a maximum birth spin-period of $\simeq 10 \, {\rm ms}$ (equation~\ref{eq:P0_max}).
Successful jets break typically out of the expanding SN ejecta on timescales of hours following the explosion (equation~\ref{eq:tbo}).

\item Observational signatures of off-axis jets associated with SLSNe are explored in \S\ref{subsec:JetSignature}.
Break-out of a transrelativistic `shocked-jet' cocoon component can produce $\sim$hr long $\sim 10^{44} \, {\rm erg \, s}^{-1}$ UV flares (\S\ref{sec:cocoon}). We estimate that $5-100$ such events may be detectable by ULTRASAT per year (equation~\ref{eq:Ndot_ULTRASAT}).
In contrast to jet break-out from a stationary stellar progenitor, we find that the Newtonian `shocked-star' cocoon component cannot overtake the expanding SN ejecta, and is therefore unobservable.

\item We have proposed a novel signature of long-duration jets --- emission from a thermal wind driven off the jet-ejecta interface (Fig.~\ref{fig:Cartoon}; \S\ref{sec:jetwind}).
We hypothesize that some, hitherto unspecified, process dissipates jet-power at this interface, driving such a wind.
The dissipation mechanism and resulting wind properties should be constrained by future high-resolution numerical studies.
Wind emission lasts for the duration of jet activity ($\sim$days for SLSNe engines) and can enhance optical SN emission, producing early light-curve maxima consistent with observed SLSNe (Fig.~\ref{fig:lightcurve}).

\item Our model generically predicts radio afterglow emission from off-axis SLSNe jets. Although radio follow-up observations have so far yielded only upper limit constraints, we show that these are consistent with expected $\gtrsim 30^\circ$ off-axis observer viewing angles (\S\ref{sec:RadioAfterglow}).  We strongly encourage a more systematic study of the late-time radio properties of SLSNe, a result with potential implications also for the origin of fast radio bursts \citep{Metzger+17,Nicholl+17b}.

\item Finally, we have illustrated the diversity of transients which can arise from the collapse of rapidly rotating stellar cores (Fig.~\ref{fig:ParameterSpace}), 
and explored statistical properties of the population of jetted transients (Fig.~\ref{fig:jets}).

\end{enumerate}

\section*{Acknowledgments}

BM and BDM gratefully acknowledge support from the NSF grant AST-1410950 and NASA grants NNX15AR47G and NNX16AB30G.  
TAT thanks Chris Kochanek for discussions.

\bibliographystyle{mnras}
\bibliography{ms}

\appendix
\section{Self-Similar Jet Solutions} \label{sec:Appendix_Jet}
Here we extend the results of \cite{Bromberg+11} to the case of a non-relativistic collimated jet propagating within a {\it time-dependent} power-law density profile
\begin{equation}
\rho \propto r^{-\delta} t^{-\beta} ,
\end{equation}
and a power-law jet luminosity
\begin{equation}
L_{\rm j} \propto t^{-\ell} .
\end{equation}
As pointed out in \S\ref{subsec:BreakoutConditions} and shown later in this Appendix, there is no solution for $\ell \geq 1$. Therefore, one should keep in mind $\ell \approx 0$ as the canonical case, appropriate for the early ($t \lesssim t_{\rm e}$) stages of the central-engine evolution, where the injected power is approximately temporally constant.

The, isobaric, cocoon pressure can be expressed as
\begin{equation} \label{Appendix:eq:P_c}
P_{\rm c} = \frac{E_{\rm c}}{3 V_{\rm c}} 
\approx \frac{\int_0^t L_{\rm j}(t^\prime) dt^\prime}{3 \int_0^{z_{\rm h}} \pi \left( \int_0^t v_{\rm c}(z_{\rm h},t^\prime) dt^\prime \right)^2 dz^\prime} ,
\end{equation}
where we have neglected terms of order $v_{\rm h}/c$, and where 
\begin{equation}
v_{\rm c}(z_{\rm h},t) = \sqrt{P_{\rm c}(t) \big/ \varrho \rho(z_{\rm h},t)}
, ~~~ \varrho = {3}/{(3-\delta)}
\end{equation}
is the cocoon's average (lateral) expansion velocity \citep{Bromberg+11}.

Assuming a power-law temporal evolution of the cocoon pressure and jet-head position, $P_{\rm c} \propto t^a$, $z_{\rm h} \propto t^b$, the cocoon volume can be integrated to obtain
\begin{align}
V_{\rm c}(t) 
&= \frac{\pi}{\varrho A^2} \times {\frac{P_{\rm c}(t) z_{\rm h}(t) t^2}{\rho\left(z_{\rm h}(t),t\right)}} ,
\end{align}
where
\begin{equation}
A = \frac{2 + \beta + a + b \delta }{2} .
\end{equation}
This, via equation~(\ref{Appendix:eq:P_c}), yields the cocoon pressure
\begin{equation} \label{Appendix:eq:P_c2}
P_{\rm c} = \sqrt{\frac{\varrho A^2 \zeta_{z}}{3 \pi (1-\ell)}} \times \sqrt{\frac{L_{\rm j}(t) \rho\left(z_{\rm h}(t),t\right)}{z_{\rm h}(t)^2 / v_{\rm h}(t)}} .
\end{equation}
Here we have substituted $z_{\rm h}(t) = \zeta_z v_{\rm h}(t) t$, which defines an integration constant $\zeta_z$ to be specified later.

\cite{Bromberg+11} find that the jet cross-section in the collimated regime is $A_{\rm j} = L_{\rm j} / 4 \gamma_{\rm j}^2 c P_{\rm c}$, which allows us to find the dimensionless jet parameter $\tilde{L} = / A_{\rm j} \rho c^3$ and hence the jet-head velocity
\begin{equation}
\frac{v_{\rm h}}{c} \approx \tilde{L}^{1/2} = \left(\frac{4 \gamma_{\rm j}^2 P_{\rm c}}{\rho c^2}\right)^{1/2} .
\end{equation}
Substituting $P_{\rm c}$ which itself depends on the jet-head velocity, we obtain
\begin{align}
v_{\rm h}(t) = \left[ \zeta_{\tilde{L}} \times \frac{L_{\rm j}(t) \gamma_{\rm j}^4}{\rho\left(z_{\rm h}(t),t\right) z_{\rm h}(t)^2} \right]^{1/3}  , ~~~
\zeta_{\tilde{L}} \equiv \frac{16 \varrho A^2 \zeta_z}{3 \pi (1-\ell)}
.
\end{align}

Identifying $v_{\rm h} = dz_{\rm h}/dt$ and integrating yields the power-law solution for the jet-head propagation
\begin{equation}
z_{\rm h}(t) \propto t^{b} \, , ~~~ b = \frac{3+\beta-\ell}{5-\delta} 
\end{equation}
This along with equation~(\ref{Appendix:eq:P_c2}) yields the closure relation
\begin{equation}
a = -\frac{1 + \beta + \ell + b(\delta + 1)}{2} = -\frac{4+\delta-\delta \ell + 2 \ell + 3\beta}{5-\delta} ,
\end{equation}
so that
\begin{align} \label{eq:zeta_Ltilde}
\zeta_{\tilde{L}} = \frac{16 (3 + \beta - \ell)}{\pi (3-\delta)(5-\delta)(1-\ell)} ,
\end{align}
and $\zeta_z = 1/b$, i.e.
\begin{equation} \label{eq:zeta_z}
\zeta_z = \frac{5-\delta}{3+\beta-\ell}  .
\end{equation}
Two useful cases of equations~(\ref{eq:zeta_Ltilde},\ref{eq:zeta_z}) are for: (a) homologous expansion, $\beta = 3-\delta$, as suitable for the expanding SN ejecta density profile (equation~\ref{eq:rho_ejecta}) and (b) stationary medium, $\beta=0$, which describes the initial stellar progenitor profile (equation~\ref{eq:rho_stellar}; here ``$\delta$'' should be identified with $w$).
In either case, the jet can only break-out successfully before the engine turn-off time, $t_{\rm e}$, so that we take $\ell=0$ (i.e. constant engine power) in equations~(\ref{eq:zeta_Ltilde},\ref{eq:zeta_z}) as the fiducial case.


\section{Jet-energized-wind model} \label{sec:Appendix_B}
The approximate light-curve produced by the jet-energized wind can be found by integrating a simple variation to the standard one-zone SN equations on the wind internal energy $E(t)$, velocity $v(t)$ ($=v_{\rm w}$ in this scenario), outer radius $R(t)$, and total mass $M(t)$ \citep[e.g.][]{Metzger+15}:
\begin{align}
&\frac{dE}{dt} = -E\frac{v}{R} -E\frac{4 \pi  c R}{3 \kappa M} + \dot{E}_{\rm w} ; \label{eq:dEdt_wind} \\
&{dv}/{dt} = {E M}/{R} ; ~~~~~
{dR}/{dt} = v ; ~~~~~
{dM}/{dt} = \dot{M}_{\rm w} . \label{eq:dxdt_wind_part2}
\end{align}
We assume that that the wind velocity is roughly constant $v \approx v_{\rm w}$ and there is no appreciable initial mass (and hence energy) in the outflow, i.e. $t_{\dot{M}} < t_{\rm exp,0}$, where $t_{\rm exp,0}=R_0/v_0$ is the initial expansion timescale 
and $t_{\dot{M}} = M_0 / \dot{M}_{\rm w}$ is mass-loss timescale. In this case, the peak emission time (set by equating the diffusion timescale to the expansion timescale) is given by equation~(\ref{eq:t_pk_wind}), $t_{\rm pk} = 3 \kappa \dot{M}_{\rm w} / 4 \pi c v_{\rm w}$, and the energy equation (\ref{eq:dEdt_wind}) can be recast in terms of the radiated luminosity $L$ (second term on the RHS of equation~\ref{eq:dEdt_wind}),
\begin{equation} \label{eq:L_wind_Appendix}
\frac{dL(t)}{dt} = -L(t) \frac{t_{\rm exp,0} + t_{\rm pk} + t}{t_{\rm pk} ( t_{\dot{M}} + t )} + \dot{E}_{\rm w}(t) \frac{t_{\rm exp,0} + t}{t_{\rm pk} ( t_{\dot{M}} + t )} .
\end{equation}

Note that the time $t$ in all equations in this appendix is measured with respect to the onset of the jet-energized wind, i.e. $t=0$ here is equivalent to $t=t_{\rm bo}$ in previous sections (where time was measured with respect to the SN explosion epoch).
In the limits $t \ll t_{\dot{M}}, t_{\rm exp,0} \ll t_{\rm pk}$, the approximate solution to equation~(\ref{eq:L_wind_Appendix}), assuming also $\dot{E}_{\rm w} \sim {\rm constant}$ (i.e. $t \ll t_{\rm e}$), is 
\begin{equation}
L(t) \approx \dot{E}_{\rm w} \frac{t_{\rm exp,0}}{t_{\rm pk}} .
\end{equation}
Next, in the limits $t_{\dot{M}}, t_{\rm exp,0} \ll t \ll t_{\rm pk}$, equation~(\ref{eq:L_wind_Appendix}) reduces to 
\begin{equation}
\frac{dL}{dt} \approx - \frac{L}{t} + \frac{\dot{E}_{\rm w}}{t_{\rm pk}}
\end{equation}
whose solution (again, for constant wind power) is
\begin{equation}
L(t) \approx \dot{E}_{\rm w} \left( \frac{t-t_{\rm exp,0}}{2t_{\rm pk}} + \frac{t_{\rm exp,o}^2}{t_{\rm pk} t} \right) .
\end{equation}
Finally, in the limit where $t_{\dot{M}}, t_{\rm exp,0} \ll t_{\rm pk} \ll t$, we find
\begin{equation}
L(t) \approx \dot{E}_{\rm w} \left( 1 - \frac{1}{2} e^{1-t/t_{\rm pk}} \right)
\end{equation}

Equations~(\ref{eq:dEdt_wind},\ref{eq:dxdt_wind_part2}) should generally be integrated numerically to produce model light-curves (this is the procedure adopted in creating Fig.~\ref{fig:lightcurve} for example), but we find that a convenient analytic functional form which schematically tracks the bolometric light-curve in the limits described above is
\begin{equation}
L(t) \sim \dot{E}_{\rm w}(t) \times \left[ 1 - e^{-(t+t_{\rm exp,0})/t_{\rm pk}} \right] .
\end{equation}
This analytic form may be more easily implemented in fitting procedures, and for the most part produces similar light-curves to the full calculation, but should non-the-less be used with caution.

\bsp	
\label{lastpage}
\end{document}